\documentclass{aa}
\usepackage{natbib}
\usepackage{amssymb,amsmath,amsfonts}
\usepackage{epsfig}
\usepackage{mathptmx}
\usepackage{longtable}
\usepackage{graphicx}
\usepackage{comment}
\usepackage{color}
\usepackage[caption=false]{subfig}
\usepackage{animate}
\usepackage{bigfoot}
\usepackage{rotating}
\usepackage{pdflscape}
\usepackage[draft]{hyperref}

\begin{document} 

\title{The LOFAR Two-metre Sky Survey -- I. Survey Description and Preliminary Data Release}

\authorrunning{Shimwell et~al.}
\titlerunning{The LOFAR Two-metre Sky Survey}
\author{T. W. Shimwell$^{1}$\thanks{E-mail: shimwell@strw.leidenuniv.nl}, 
H. J. A. R\"{o}ttgering$^{1}$, 
P. N. Best$^{2}$, 
W. L. Williams$^{3}$,
T.J. Dijkema$^{4}$, 
F. de Gasperin$^{1}$, 
M. J. Hardcastle$^{3}$, 
G. H. Heald$^{5,6}$,  
D. N. Hoang$^{1}$, 
A. Horneffer$^{7}$, 
H. Intema$^{1}$,  
E. K. Mahony$^{4,8,9}$, 
S. Mandal$^{1}$,
A. P. Mechev$^{1}$, 
L. Morabito$^{1}$,
J. B. R. Oonk$^{1,4}$,  
D. Rafferty$^{10}$,
E. Retana-Montenegro$^{1}$, 
J. Sabater$^{2}$,  
C. Tasse$^{11,12}$,  
R. J. van Weeren$^{13}$, 
M. Br\"uggen$^{10}$, 
G. Brunetti$^{14}$, 
K. T. Chy\.zy$^{15}$,
J. E. Conway$^{16}$,  
M. Haverkorn$^{17}$, 
N. Jackson$^{18}$, 
M. J. Jarvis$^{19,20}$,  
J. P. McKean$^{4,6}$,
G. K. Miley$^{1}$,  
R. Morganti$^{4,6}$,  
 G. J. White$^{21,22}$, 
 M. W. Wise$^{4,23}$,
I. M. van Bemmel$^{24}$,  
R. Beck$^{7}$, 
M. Brienza$^{4,6}$, 
A. Bonafede$^{10}$,  
G. Calistro Rivera$^{1}$,  
R. Cassano$^{14}$,
A. O. Clarke$^{18}$,
D. Cseh$^{17}$,
A. Deller$^{4}$,
A. Drabent$^{25}$,  
W. van Driel$^{11,26}$,
D. Engels$^{10}$, 
H. Falcke$^{4,17}$, 
C. Ferrari$^{27}$,
S. Fr\"{o}hlich$^{28}$, 
M. A. Garrett$^{4}$,
J. J. Harwood$^{4}$,  
V. Heesen$^{29}$, 
M. Hoeft$^{24}$,  
C. Horellou$^{16}$,
F. P. Israel$^{1}$,
A. D. Kapi\'nska$^{9,30,31}$,
M. Kunert-Bajraszewska$^{32}$,    
D. J. McKay$^{33,34}$,
N. R. Mohan$^{35}$,
E. Orr\'u$^{4}$,
R. F. Pizzo$^{4}$,  
I. Prandoni$^{14}$, 
D. J. Schwarz$^{36}$, 	
A. Shulevski$^{4}$,
M. Sipior$^{4}$,
D. J. B. Smith$^{3}$,  
S. S. Sridhar$^{4,6}$,
M. Steinmetz$^{37}$,
A. Stroe$^{38}$,  
E. Varenius$^{16}$,
P. P. van der Werf$^{1}$, 
J. A. Zensus$^{7}$,
J. T. L. Zwart$^{20,39}$ \\
(Affiliations can be found after the references)}
\institute{}
\date{Accepted ---; received ---; in original form \today}

\abstract{
\noindent
The LOFAR Two-metre Sky Survey (LoTSS) is a deep 120-168\,MHz imaging survey that will eventually cover the entire Northern sky. Each of the 3170 pointings will be observed for 8\,hrs, which, at most declinations, is sufficient to produce $\sim$5$\arcsec$ resolution images with a sensitivity of $\sim$100\,$\mu$Jy/beam and accomplish the main scientific aims of the survey which are to explore the formation and evolution of massive black holes, galaxies, clusters of galaxies and large-scale structure. Due to the compact core and long baselines of LOFAR, the images provide excellent sensitivity to both highly extended and compact emission. For legacy value, the data are archived at high spectral and time resolution to facilitate subarcsecond imaging and spectral line studies. In this paper we provide an overview of the LoTSS. We outline the survey strategy, the observational status, the current calibration techniques, a preliminary data release, and the anticipated scientific impact. The preliminary images that we have released were created using a fully-automated but direction-independent calibration strategy and are significantly more sensitive than those produced by any existing large-area low-frequency survey. In excess of 44,000 sources are detected in the images that have a resolution of 25$\arcsec$,  typical noise levels of less than 0.5\,mJy/beam, and cover an area of over 350 square degrees in the region of the HETDEX Spring Field (right ascension 10h45m00s to 15h30m00s and declination 45$^\circ$00$\arcmin$00$\arcsec$ to 57$^\circ$00$\arcmin$00$\arcsec$).}

\maketitle

\section{Introduction}

Performing increasingly sensitive surveys is a fundamental endeavour of astronomy. Over the past 60 years, the depth, fidelity, and resolution of radio surveys has continuously improved. However, new, upgraded and planned instruments are capable of revolutionising this area of research. The International Low-Frequency Array (LOFAR; \citealt{vanHaarlem_2013}) is one such instrument. LOFAR offers a transformational increase in radio survey speed compared to existing radio telescopes. It also opens up a poorly explored low-frequency region of the electromagnetic spectrum. An important goal that has driven the development of LOFAR since its inception is to conduct wide and deep surveys.  The LOFAR Surveys Key Science Project (PI: R\"{o}ttgering) is conducting a survey with three tiers of observations: Tier-1 is the widest tier and includes low-band antenna (LBA) and high-band antenna (HBA) observations across the whole 2$\pi$ steradians of the Northern sky; deeper Tier-2 and Tier-3 observations are focussing on smaller areas with high-quality multi-wavelength datasets.

Here we focus on the {ongoing} LOFAR HBA 120-168\,MHz Tier-1 survey, hereafter referred to as the LOFAR Two-metre Sky Survey (LoTSS). {This is the second northern hemisphere survey that will be conducted with the LOFAR HBA and is significantly deeper than the first, the Multifrequency Snapshot Sky Survey (MSSS; \citealt{Heald_2015}). MSSS was primarily conducted as a commissioning project for LOFAR and a testbed for large-scale imaging projects, whereas LoTSS will probe a new parameter space. LoTSS is a long term project but over 2,000 square degrees of the northern sky have already been observed and additional data are continuously being taken. }

{The main scientific motivations for LoTSS} are to explore the formation and evolution of massive black holes, galaxies, clusters of galaxies and large-scale structure. More specifically, the survey was initially designed to detect: 100 radio galaxies at $z > 6$ (based on the predicted source populations of \citealt{Wilman_2008}); and diffuse radio emission associated with the intra-cluster medium of 100 galaxy clusters at $z>0.6$ (\citealt{Ensslin_2012} and \citealt{Cassano_2010}); along with up to $3\times10^7$ other radio sources. In addition, the survey had to meet practical requirements such as high efficiency, manageable data rates with sufficient time and frequency resolution, a workable data processing strategy, good $uv$ plane coverage with sensitivity to a wide range of angular scales, and a feasible total duration. These criterion resulted in the ambitious observational aims of producing high-fidelity 150\,MHz images of the entire Northern sky that have a resolution of $\sim5\arcsec$ and sensitivity of $\sim100\,\mu$Jy/beam at most declinations (equivalent to a depth of $\sim20\,\mu$Jy/beam at 1.4\,GHz for a typical synchrotron radio source of spectral index $\alpha \sim -0.7$, where the radio flux density $S_\nu \propto \nu^{\alpha}$). 

Besides the primary objectives there are many other important science factors that have further motivated the LoTSS. The survey will significantly increase the known samples of young and old AGN, including giant, dying and relic sources, allowing detailed studies of the physics of AGN. It will also detect millions of AGN out to the highest redshifts (\citealt{Wilman_2008}), including obscured AGN, radiatively-inefficient AGN, and `radio-quiet' AGN, and thus allow statistical studies of the evolution of the properties of different classes of AGN over cosmic time (e.g. \citealt{Best_2014}). The sensitive images of the steep spectrum radio emission from local galaxy clusters, and the expected detection of hundreds of galaxy clusters out to moderate redshifts, will transform our  knowledge of magnetic fields and particle acceleration mechanisms in clusters (e.g. \citealt{Cassano_2010}). {Hundreds of thousands of star-forming galaxies will be detected, primarily at lower redshifts but extending out to $z\ga1$. These} will be used to distinguish between various models that describe the correlation between the low frequency radio continuum and the far-infrared emission and the variation of this correlation with galaxy properties (e.g. \citealt{Hardcastle_2016} and \citealt{Smith_2014}). {They will also trace the cosmic star-formation rate density in a manner unaffected by the biases of dust obscuration or source confusion (e.g. \citealt{Jarvis_2015}).}  {The survey images, in combination with other datasets, will be used to measure cosmological parameters, including tests of alternative theories of gravity, and using the Integrated Sachs-Wolfe effect to constrain the nature of dark energy (e.g. \citealt{Raccanelli_2012},  \citealt{Jarvis_2015} and \citealt{Schwarz_2015}).}  Detailed maps of nearby galaxies will be used for studies of cosmic ray diffusion and magnetic fields. The shortest LOFAR baselines (less than 50\,m) allow for degree scale emission to be accurately recovered, and the number of well imaged supernova remnants and \ion{H}{II} regions will be increased by an order of magnitude to forward studies of the interstellar medium and star formation. Galactic synchrotron emission mapping will provide new information about the strength and topology of the large-scale galactic magnetic field (\citealt{Iacobelli_2013}). 

{In addition, the survey datasets will be used for a range of other projects. The low-frequency polarisation maps will be used by the Magnetism Key Science project to measure the Faraday spectra of sources (\citealt{Beck_2013}).  } The high spectral resolution makes it possible to investigate the physics of the cold, neutral medium in galaxies and its role in galaxy evolution by means of radio recombination lines (e.g. \citealt{Oonk_2014} and \citealt{Morabito_2014}). The wide area coverage will allow for tight constraints on the population of transient  sources and the exploration of new parameter space will open up the possibility of serendipitous discoveries. {The eventual exploitation of international baselines will facilitate science that requires subarcsecond resolution.} For example, it will allow us to access a regime in which AGN and star-forming galaxies can be accurately distinguished by morphology (e.g. \citealt{Muxlow_2005}) {, and because} of the large number of detected sources, we will also be able to discover rare objects such as strongly lensed radio sources {which can} yield constraints on galaxy evolution (e.g \citealt{Sonnenfeld_2015}) and the distribution of dark matter substructure (see  \citealt{Jackson_2013} and references within). 

\begin{figure}   \centering
   \includegraphics[width=\linewidth]{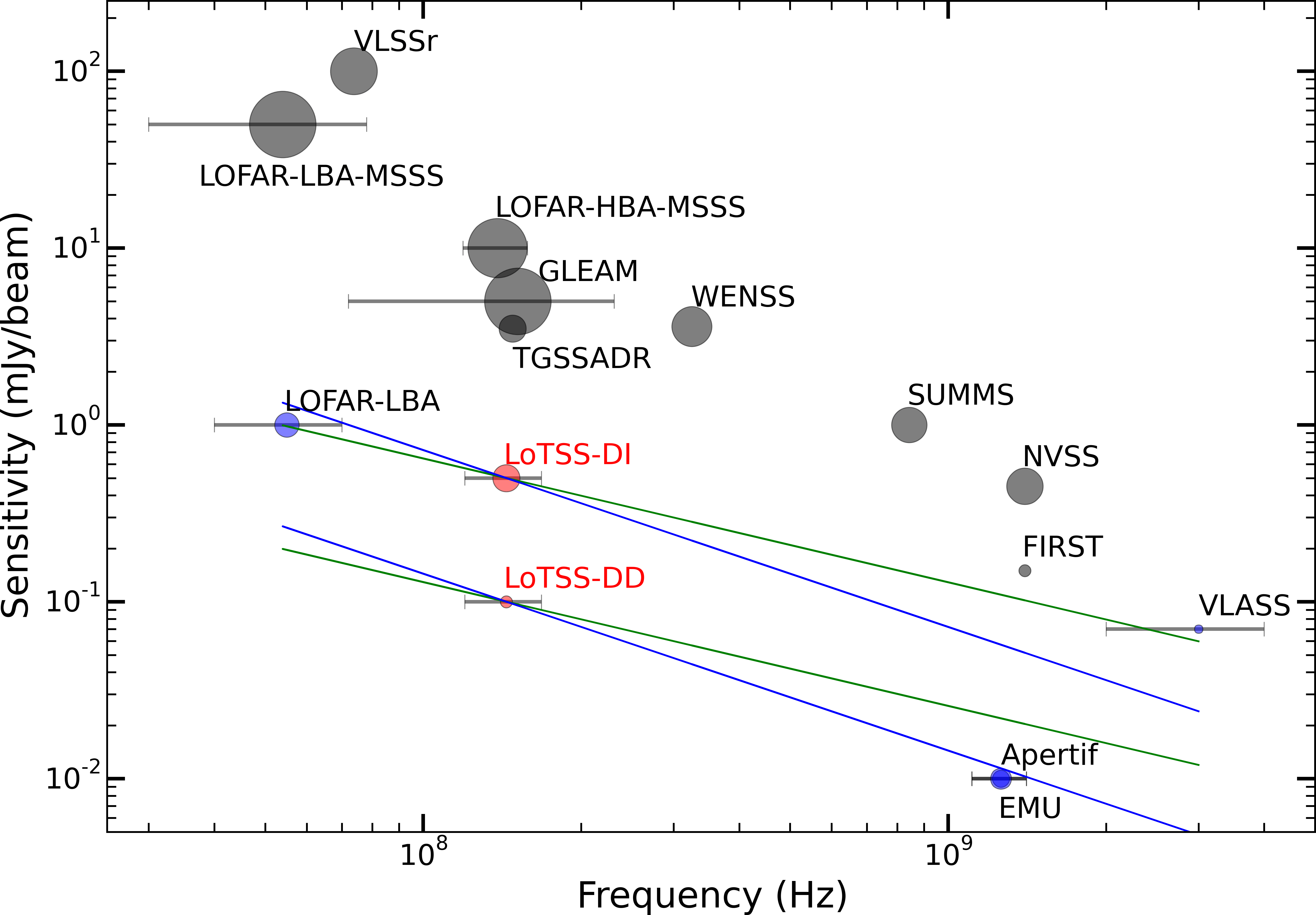}
   \caption{A summary of the sensitivity, frequency and resolution of a selection of recent and planned large-area radio surveys (see also Table \ref{tab:lowfreq-survey-sensitivity}). The size of the markers is proportional to the square root of the survey resolution. Grey, blue and red markers show the ongoing/completed surveys, forthcoming surveys, and the LOFAR HBA surveys respectively. {The horizontal lines show the frequency coverage for surveys with large fractional bandwidths ($>0.2)$. The green sloping lines} show the sensitivity that is equivalent to that achieved in the LoTSS direction-dependent (DD) calibrated and direction-independent (DI) images for typical radio sources with a spectral index $\sim-0.7$ . Similarly, the {blue sloping lines} show the equivalent sensitivity to steep spectrum sources with a spectral index $\sim-1.0$.}
   \label{fig:comp_sensitivity}
\end{figure}

The long integration time on each survey grid pointing that can be afforded due to the wide field of view of the HBA stations, together with the extensive range of baseline lengths in the array, allow the LoTSS to probe a combination of depth, area, resolution and sensitivity to a wide range of angular scales that has not previously been achieved in any wide-area radio survey (see Figure \ref{fig:comp_sensitivity}). For example, in comparison to other recent low-frequency surveys, such as the TIFR GMRT Sky Survey alternative data release (TGSS; \citealt{Intema_2016}), {MSSS (\citealt{Heald_2015}),} GaLactic and Extragalactic All-sky MWA (GLEAM; \citealt{Wayth_2015}) and the Very Large Array Low-frequency Sky Survey Redux (VLSSr; \citealt{Lane_2014}), the 120-168\,MHz LoTSS will be at least a factor of 50-1000 more sensitive and 5-30 times higher in resolution (see Table \ref{tab:lowfreq-survey-sensitivity}). 

{In} comparison to higher frequencies the LoTSS will match the high resolution achieved by Faint Images of the Radio Sky at Twenty-Centimeters (FIRST; \citealt{Becker_1995}) but over a wider area and, for a typical radio source of spectral index $\alpha \sim-0.7$, it will be 7 times more sensitive. Similarly, the LoTSS will be 20 times more sensitive to typical radio sources than the lower resolution  NRAO VLA Sky Survey (NVSS; \citealt{Condon_1998}) and the dense core of LOFAR provides a large improvement in surface brightness sensitivity. There are other large upcoming radio surveys that are mutually complementary with the LoTSS.  For example, the LOFAR HBA and LBA sky surveys will be exceptionally sensitive to steep spectrum ($\alpha \leq -1$) objects. By comparison, the Evolutionary Map of the Universe (EMU; \citealt{Norris_2011}) and APERture Tile In Focus (Apertif; \citealt{Rottgering_2011}) 1.4\,GHz surveys, whilst at lower resolution, aim to reach a depth of  $\sim10\,\mu$Jy/beam (corresponding to 50\,$\mu$Jy/beam at 150\,MHz for $\alpha \sim -0.7$)  and will offer improved sensitivity to typical or flatter spectrum radio emission. Meanwhile, the 1-3\,GHz VLA Sky Survey (VLASS\footnote{https://science.nrao.edu/science/surveys/vlass}), will not survey as deeply, but will provide images with 2.5$\arcsec$ resolution to pinpoint the precise location of sources. 

In this publication, we describe the LoTSS strategy, and the current calibration and imaging techniques. We also release preliminary 120-168\,MHz images and catalogues of over 350 square degrees from right ascension of 10h45m00s to 15h30m00s and declination 45$^\circ$00$\arcmin$00$\arcsec$ to 57$^\circ$00$\arcmin$00$\arcsec$ which is in the region of the Hobby-Eberly Telescope Dark Energy Experiment (HETDEX) Spring Field (\citealt{Hill_2008}). This field was targeted as it is a large contiguous area at high elevation for LOFAR,  whilst having a large overlap with the Sloan Digital Sky Survey (SDSS; \citealt{York_2000}) imaging and spectroscopic data. Importantly, it also paves the way for using HETDEX data to provide emission-line redshifts for the LOFAR sources and prepares for the WEAVE-LOFAR\footnote{http://www.ing.iac.es/weave/weavelofar/} survey which will measure spectra of more than $10^6$ LOFAR-selected sources (\citealt{Smith_2015}). The region was also chosen because HETDEX is a unique survey that is very well matched to the key science questions that the LOFAR surveys aims to address. In particular, the ability to obtain [\ion{O}{II}] redshifts up to $z\sim 0.5$ is well matched to the LOFAR goal of tracking the star-formation rate density using radio continuum {observations. Furthermore,} the main science goal of HETDEX is to obtain emission line redshifts using Ly$\alpha$ at $1.9<z<3.5$, which is around the peak in the space density of powerful AGN as well as the peak of the star formation rate and the merger rate of galaxies (\citealt{Jarvis_2000}, \citealt{Rigby_2015}, \citealt{Madau_2014} and \citealt{Conselice_2014}), and will thus help to provide the necessary data for a full census of radio sources over this cosmic epoch. The LOFAR data can help the HETDEX survey to distinguish between low-redshift [\ion{O}{II}] and high-redshift Ly$\alpha$ emitters, e.g. using the Bayesian framework set out in \cite{Leung_2015}. 

The greatest challenge we face in reaching the observational aims of the LoTSS is to routinely perform an accurate, robust, and efficient calibration of large datasets to minimise the direction-dependent effects that severely limit the image quality.  This complex direction-dependent calibration procedure, which corrects for the varying ionospheric conditions (e.g. \citealt{Mevius_2016}) and errors in the beam models, is crucial to create high-fidelity images at full resolution and sensitivity. Several approaches are being developed to minimise these direction-dependent effects (e.g. \citealt{Tasse_2014} and \citealt{Yatawatta_2015}), including the facet calibration procedure (\citealt{vanWeeren_2015a} and \citealt{Williams_2015}). This procedure has already been successfully applied to several fields to produce high-resolution images with high fidelity and a sensitivity approaching the thermal noise (\citealt{Williams_2015}, \citealt{vanWeeren_2015b}, \citealt{Shimwell_2016} and \citealt{Hardcastle_2016}). 

{A} direction-dependent calibration technique will be used to calibrate all LoTSS data in the future to produce images that meet our observational aims, {but} the exact procedure is still being finalised. Therefore, for this publication, we simply demonstrate that we can achieve these ambitious imaging aims by performing a direction-dependent calibration of a single randomly chosen field to produce an 120-168\,MHz image with $4.8\arcsec \times7.9\arcsec$ resolution and 100\,$\mu$Jy/beam sensitivity. However, our large data release consists of preliminary images and catalogues that were instead created with a rapid and automated direction-independent calibration of the 63 HBA pointings that cover over 350 square degrees in the region of the HETDEX Spring Field. Although ionospheric and beam effects do hinder the image fidelity of these preliminary images, we are able to image data from baselines shorter than 12\,k$\lambda$ to produce $25\arcsec$ resolution images that typically have a noise level of 200-500\,$\mu$Jy/beam away from bright sources. Such sensitive, low-frequency images have not previously been produced over such a wide area and are sufficient to accomplish many of the scientific objectives of the survey (see \citealt{Brienza_2016}, \citealt{Harwood_2016}, \citealt{Heesen_2016}, \citealt{Mahony_2016}, \citealt{Shulevski_2015a} and \citealt{Shulevski_2015b} for examples). 

The outline of this paper is as follows. In Section \ref{sec:survey-strategy}, we describe the survey strategy including the choice of observing mode, frequency coverage, dwell time, tiling, and the data that are archived. The status of the observing programme for the LoTSS is summarised in Section \ref{sec:obsstatus}. In Sections \ref{sec:datareduction}, \ref{sec:imagequality}, \ref{sec:mosaicing} and \ref{sec:sourcecatalogues} we describe the calibration techniques, imaging procedure, image quality and source cataloguing that we have used for this preliminary data release. The data release itself is summarised in Section \ref{sec:publicddatarelease}. In Section \ref{sec:facet-calibration} we provide an example of the improvement in image fidelity, sensitivity and resolution that will be achieved once direction-dependent calibration has been performed on our datasets. Section \ref{sec:scientificpotential} provides a brief overview of the scientific potential of the LoTSS data before we summarise in Section \ref{sec:summary}.

\begin{table*}
\caption{A summary of recent large area low-frequency surveys (see also Figure \ref{fig:comp_sensitivity}). We have attempted to provide a fair comparison of sensitivities and resolutions but we note that both the sensitivity and resolution achieved varies within a given survey.}
 \centering
 \label{tab:lowfreq-survey-sensitivity}
\begin{tabular}{lcccccc}
\hline 
Survey & Resolution & Noise & Frequency  & Area \\
                            & ($\arcsec$) & (mJy/beam) & (MHz) & \\ \hline
GLEAM (\citealt{Wayth_2015}) & 150  & 5 & 72--231 & $\delta<+25^\circ$ \\
MSSS-HBA (\citealt{Heald_2015}) & 120 & 10           & 119--158 & $\delta>0^\circ$ \\
MSSS-LBA (\citealt{Heald_2015}) & 150 & 50           & 30--78 & $\delta>0^\circ$ \\
TGSS ADR (\citealt{Intema_2016}) & 25  & 3.5            & 140--156 & $\delta>-53^\circ$ \\
LoTSS direction-dependent  & 5 & 0.1 & 120--168 & $\delta>0^\circ$ \\ 
LoTSS direction-independent (this paper)  & 25 & 0.5 & 120--168 & HETDEX Spring Field \\ 
VLSSr (\citealt{Lane_2014}) & 75     & 100           &  73--74.6 & $\delta>-30^\circ$ \\ \hline
 \end{tabular}
\end{table*}

\section{Survey strategy}
\label{sec:survey-strategy}
Prior to routinely undertaking observations for the large-scale LoTSS, the array configuration, integration time, frequency coverage, and tiling strategy were chosen. The main aim of the LOFAR HBA survey is to observe the entire Northern sky and achieve a resolution of 5$\arcsec$ and a sensitivity of $\sim$100\,$\mu$Jy/beam at most declinations. In this section we outline the strategy we have adopted to efficiently conduct a survey that can accomplish this goal which is summarised in Table \ref{tab:LOFAR-obs}. In choosing our observing setup we bore in mind that, for legacy value, the archived data should be able to facilitate as much science as possible. The archived data should be capable of exploiting the facts that LOFAR has a native spectral resolution suitable for spectral line studies and, while the majority of LOFAR stations are in the Netherlands, at the time the data presented here were taken, there were also international stations in Germany, France, Sweden and the UK that provide baselines up to 1300\,km. The array has been further extended during 2016 to increase the maximum baseline length to 1600\,km with three new stations in Poland, and a station in Ireland is currently under construction. These international stations will allow HBA imaging at resolutions of $\sim0.3\arcsec$. Imaging at the full resolution provided by the international stations has been shown to be possible for individual targets (e.g. \citealt{Varenius_2015} with the HBA and \citealt{Morabito_2016} with the LBA), reaching sensitivities of 150\,$\mu$Jy/beam for the HBA. Accordingly, international stations are present in the LoTSS datasets, although these data are not yet routinely imaged as part of the Survey programme. Such routine imaging will require further work on identification of calibrator sources with significant compact structure, which is currently being undertaken by the LBCS project (\citealt{Moldon_2015} and \citealt{Jackson_2016}). It will also require further work on the calibration and understanding of ionospheric effects, which is currently under way (e.g. \citealt{Mevius_2016}).

\subsection{Observing mode}

LOFAR can observe with several different configurations of the HBA tiles, which are described in \citealt{vanHaarlem_2013} and on the observatory's webpage\footnote{https://www.astron.nl/radio-observatory/astronomers/technical-information/lofar-technical-information}. The configurations that affect the core stations are: \textsc{hba\_zero} or \textsc{hba\_one}, which make use of only one of the two sub-stations in each core station; \textsc{hba\_dual}, which  correlates the signal from each sub-station in each core station separately; and  \textsc{hba\_joined}, where the two sub-stations in each core station act as a single station which results in different beam shapes for different stations. For each configuration the number of tiles used on a remote station can also be selected to be either the inner 24 tiles (to match the core station sub-stations) or the full 48 tiles. At the time of writing, international stations always observe with their full 96 tiles. For the LoTSS, we decided to use  \textsc{hba\_dual\_inner}, where all stations within the Netherlands operate with 24 tiles and each sub-station in the core stations is correlated separately. This configuration was chosen because it does not  reduce the number of short baselines or suffer from additional calibration difficulties caused by non-uniform beam shapes. By discarding  24 of the 48 tiles of the remote stations, we reduce the sensitivity but gain a wider field of view. 

\subsection{Observing bandwidth and integration time}

Both the dwell time on each survey pointing and the frequency range allocated, are primarily dictated by the desired sensitivity of $\sim$100\,$\mu$Jy/beam but this must be coupled with the need for efficient observing and the desire to simplify book-keeping and scheduling. The most efficient HBA observing is performed using the 110-190\,MHz band, which has the least radio frequency interference (RFI) of the available LOFAR HBA bands. By recording data with 8-bits per sample (at the time of writing a 4-bit mode is being developed but is not yet available for observing) up to 488 195.3\,kHz wide sub-bands are available for observing. These sub-bands can be split between multiple station beams, which, for high-sensitivity, must be positioned within the HBA tile beam, which has a full width half maximum (FWHM) of 20$^\circ$ at 140\,MHz (see \citealt{vanHaarlem_2013} for a detailed description of the LOFAR beams). To achieve our target sensitivity, the entire 110-190\,MHz is not required, as the System Equivalent Flux Density (SEFD) measurements provided by \cite{vanHaarlem_2013} imply that observing for 8\,hrs with 48\,MHz of bandwidth within the 110-190\,MHz band will allow us to reach our target sensitivity of $\sim100\,\mu$Jy/beam. This is also supported by previous observations, for example: \cite{vanWeeren_2015b}  reach 93\,$\mu$Jy/beam noise with 120-181\,MHz coverage and 10\,hrs of observation; \cite{Williams_2015} obtain a sensitivity of 110\,$\mu$Jy/beam with  130-169\,MHz coverage and 8\,hrs of observation;  \cite{Shimwell_2016} reach 190\,$\mu$Jy/beam with 120-170\,MHz coverage and 8\,hrs of observation; and \cite{Hardcastle_2016} reach $100\,\mu$Jy/beam sensitivity with 126-173\,MHz coverage and 8\,hrs integration time. 

{To} increase the efficiency of the observing we use two station beams simultaneously with 48\,MHz of bandwidth allocated to each. The station beams are separated by between four and ten degrees to avoid correlated noise in the regions where the beams overlap, and the tile beam is centred midway between the two station beams to reduce the sensitivity loss. The LOFAR HBA sensitivity varies as a function of frequency due to the gain of the receiving elements (which drops off near the band edges) and the prevalence of RFI.  We choose to observe between 120\,MHz and 168\,MHz to avoid the frequencies within the 110-190\,MHz band that have the highest levels of RFI contamination or the poorest SEFD measurements. This frequency range was also chosen in an attempt to maximise the survey efficiency in terms of the number of sources detected -- observing towards the lower end of the HBA band increases the area of the field of view in proportion to $\nu^{-2}$ and enhances the brightness of sources in proportion to approximately $\nu^{-0.7}$. For simple scheduling, we aim to complete the majority of observations with a single integration. To achieve our sensitivity goals we opted to observe each pointing for 8\,hrs. Longer tracks were not practical because, similarly to other low-frequency phased arrays, the sensitivity of LOFAR decreases significantly when observing below 30 degrees in elevation. This is due to, for example, the reduced projected collecting area and the longer line of sight through the ionosphere.

\begin{figure*}   \centering
   \includegraphics[width=0.47\linewidth]{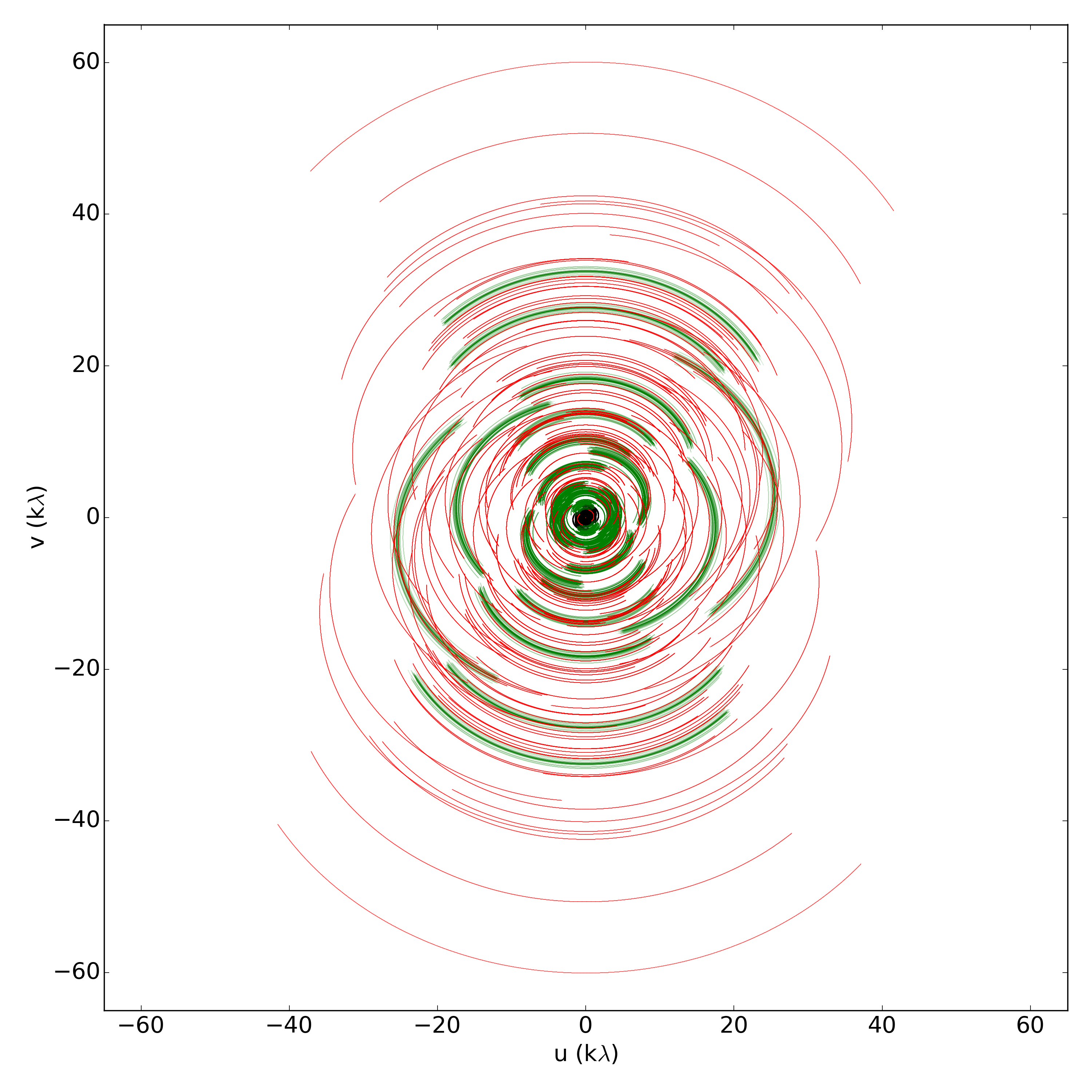} 
   \includegraphics[width=0.47\linewidth]{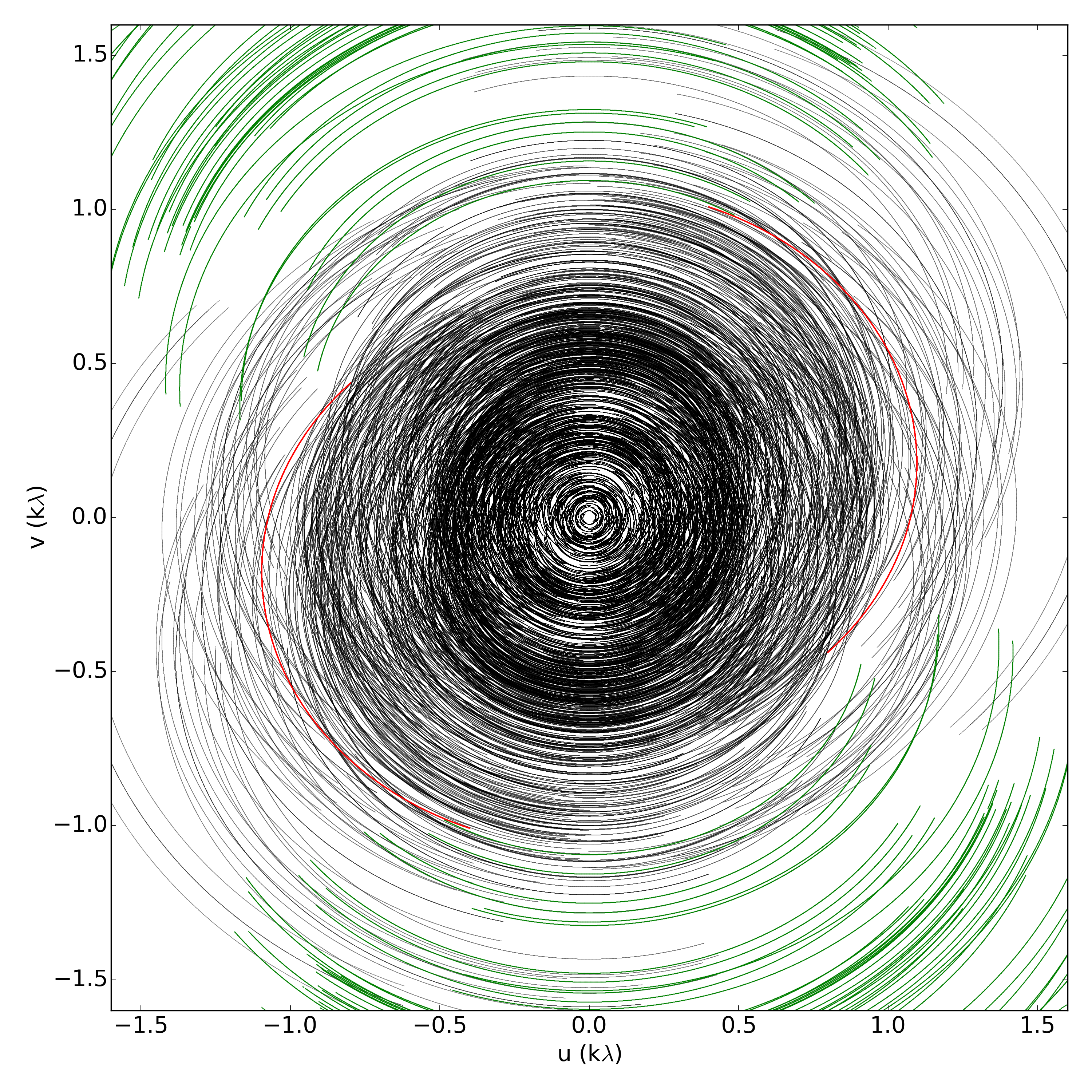} 
   \caption{The monochromatic $uv$ plane coverage of a typical 8\,hr 150\,MHz LoTSS observation around declination +55$^\circ$ excluding the international stations. On the left is the full $uv$ coverage and on the right we show the dense $uv$-coverage in the inner region of the $uv$-plane. Here the monochromatic coverage has been presented for display purposes but the full bandwidth used in each observation is 48\,MHz, which corresponds to a fractional bandwidth of $\sim$1/3, and this provides considerable additional filling of the $uv$-plane. {The $uv$ points are colour coded according to the type of stations that make up each baseline. Those containing only core stations, remote stations, or a combination of the two are shown in black, red, and green respectively.}}
   \label{fig:uv-coverage}
\end{figure*}

The typical $uv$-plane coverage of an 8\,hr LoTSS observation is shown in Figure \ref{fig:uv-coverage} (excluding the international stations). The dense core of the array produces a very high density of measurements within 2\,km which provides excellent surface brightness sensitivity. The most remote stations within the Netherlands provide baselines up to 120\,km and allow for $\sim5\arcsec$ resolution imaging. The very uneven distribution of points on the $uv$-plane implies that the naturally weighted synthesised beam when imaging with all the Dutch stations of LOFAR has high sidelobes. {However, these sidelobes} can be reduced significantly by weighting the visibilities with a more uniform weighting scheme such as the \cite{Briggs_1995} weighting scheme {and using $uv$-tapers to reduce the sharpness of cut-offs in the $uv$-plane coverage.}

\subsection{Pointing strategy}
\label{sec:pointingstrategy}
The FWHM of the LOFAR \textsc{hba\_dual\_inner} primary beam is given by 

\begin{equation}
FWHM = 1.02 \frac{\lambda}{D},
\end{equation}
where $\lambda$ is the observing wavelength and $D$ is 30.75\,m,  the diameter for the \textsc{hba\_dual\_inner}  stations (\citealt{vanHaarlem_2013}). This implies a station beam FWHM of 4.75$^\circ$ at 120\,MHz, 3.96$^\circ$ at 144\,MHz and 3.40$^\circ$ at 168\,MHz. {Nyquist sampling the LoTSS pointings at the highest observed frequency would be required to accurately reconstruct spatial scales that are similar to the primary beam size (\citealt{Cornwell_1998}) but would result in a large number of pointing centres and is not required to obtain close to uniform sensitivity across the sky. A much coarser sampling is typically used for interferometric radio surveys, for example, }at the Australia Telescope Compact Array (ATCA{\footnote{http://www.atnf.csiro.au/computing/software/miriad/}) a separation of $\rm{FWHM}/\sqrt3$ is recommended and for the Very Large Array (VLA), the NVSS survey \cite{Condon_1998} found that a separation of $\rm{FWHM/\sqrt2}$ would provide nearly uniform sensitivity coverage (the lowest sensitivity being about 90\% of the highest sensitivity) and ended up using an even coarser spacing of FWHM/1.2. These previous experiences indicate that for the highest frequency of the LOFAR HBA survey (168\,MHz) the separation between pointing centres should not exceed 2.80$^\circ$ (FWHM/1.2). However, for more uniform sensitivity the pointings should be separated by around 2.40$^\circ$ ($\rm{FWHM/\sqrt2}$). To give an indication of approximately how many pointings this requires, we find that to hexagonally tile a plane with an area equal to half the sky at 2.80$^\circ$ separation can be done with 2973 pointings while 2.40$^\circ$ separation requires 4134 pointings. The final separation we have chosen is a compromise between the time taken to observe the sky and the desired uniformity. We decided to aim for a separation of $\approx$2.58$^\circ$ which samples the sky at our lowest observed frequency close to the Nyquist criterion and approximately samples by $\rm{FWHM/\sqrt2}$ at the highest frequencies.  

Various tiling strategies have been adopted to perform large area radio surveys but many are based around the efficient hexagonal close-packed grid structure. For example, the VLA NVSS (\citealt{Condon_1998}) and FIRST (\citealt{Becker_1995}) surveys used similar strategies, adopting a hexagonal close-packed grid with a fixed right ascension separation over a certain declination range, but with a declination spacing that varied with approximately 1/cos(dec) to keep a roughly constant number of pointing centers per unit area on the sphere. The WENSS survey (\citealt{Rengelink_1997}) used a hexagonal grid with rows of constant declination throughout but altered the right ascension separation of a certain declination range. The VLSS  (\citealt{Cohen_2007}) and MSSS (\citealt{Heald_2015}) surveys, which have a much larger primary beam than the higher frequency surveys, again used an approximately hexagonal grid pattern to cover the sky but the GLEAM (\citealt{Wayth_2015}) survey, which also has a very large primary beam, used a drift scan technique over declination strips. We have adopted a slightly different scheme where our pointing positions are determined using the  \cite{Saff_1997} algorithm which attempts to uniformly distribute a large number of points over the surface of a sphere. This algorithm produces a spherical spiral distribution of pointings (see Figure \ref{fig:Tier-1-HBA-pointings}), where the pointing centres do not lie on rows of constant declination but the structure of adjacent pointing centres resembles a hexagonal close-packed grid structure. 

Using the \cite{Saff_1997} algorithm to populate the Northern hemisphere with pointings that are typically separated by 2.58$^\circ$ we have identified 3170 pointing locations which make up the LoTSS grid. The distribution of the  separation of pointing positions and the final grid for the LoTSS is shown in Figure \ref{fig:Tier-1-HBA-pointings}. We note that 42 of the first pointings to be observed were test observations for the survey and were tiled using a slightly different scheme which had a similar separation but followed rows of constant declination. Our final survey spherical spiral grid was rotated so that it best matched up with these early observations. The slight mismatch between the two strategies is apparent in Figure \ref{fig:Tier-1-HBA-pointings}.

The density of pointings in the pointing grid is approximately uniform, but it is known that at low declinations the shape of the LOFAR station beam is significantly enlarged (primarily in the north-south direction) and that the sensitivity of the array is reduced. We have not yet precisely accounted for these variations in the structure of our survey grid, but the enlargement of the station beam at lower declination will result in a larger overlap of neighbouring pointings and, while this does not eliminate the sensitivity variations with declination, it does help to reduce them. Furthermore, we have initiated a series of observations close to zero declination to observationally characterise the expected sensitivity loss.

\subsection{Archived datasets}
\label{sec:archived-data}

To facilitate both spectral line and international baseline studies, the data are not heavily averaged in either frequency or time before they are archived in the LOFAR Long Term Archive ({\footnote{http://lofar.target.rug.nl}). We have opted to store the data at 1\,s time resolution and 12.2\,kHz frequency resolution (note that some early observations have up to a factor of 4 more averaging). The effects of the time and bandwidth smearing that this averaging causes can be approximated using the equations of \cite{Bridle_1989}. The time averaging of 1\,s is such that for international station imaging at 0.5$\arcsec$ resolution, time smearing will reduce the peak brightness of sources 1$^\circ$ away from the pointing centre by 7\%. The effects of the 12.2\,kHz frequency averaging are approximately equal: at 0.5$\arcsec$ resolution and 150\,MHz the effects of bandwidth smearing will reduce the peak brightness of sources 1$^\circ$ away from the pointing centre by 8\%. 

Whilst archiving the data at such high time and frequency resolution is crucial to facilitate valuable spectral line and international baseline studies, the downside is that the data volume is very large. The dataset for each pointing is approximately 16\,TB, thus the estimated data size for the entire LoTSS is over 50\,PB. However, prior to calibrating or imaging the data for the 5$\arcsec$ resolution LoTSS, we can rapidly preprocess the data with an averaging of a factor of four in time and four in frequency. This averaging can be done because for 5$\arcsec$ imaging a time resolution of 4\,s and a frequency resolution of 48.8\,kHz is sufficient to prevent significant smearing within the LOFAR field of view. {With this averaging, at a distance of 1.85$^\circ$ from the pointing centre, which corresponds to the maximum distance at which LoTSS pointings overlap (see Figure \ref{fig:Tier-1-HBA-pointings}), we estimate a 3\% peak brightness loss due to  time averaging smearing and a 4\% peak brightness loss due to bandwidth smearing. }

\begin{table}\caption{A summary of the LoTSS survey properties. The sensitivity and noise estimates are appropriate for most observations but we note that the sensitivity may be reduced at low declination (see Section \ref{sec:pointingstrategy}).}
 \centering
 \label{tab:LOFAR-obs}
\begin{tabular}{lccc}
\hline 
Number of pointings & 3170 \\ 
Separation of pointings & 2.58$^\circ$ \\
Integration time & 8\,hrs \\
Frequency range & 120-168\,MHz\\
Array configuration & \textsc{hba\_dual\_inner}\\
Angular resolution & $\sim$5$\arcsec$\\
Sensitivity & $\sim$100\,$\mu$Jy/beam\\
Time resolution & 1\,s$^*$\\
Frequency resolution & 12.2\,kHz$^*$\\ \hline
 \end{tabular}
\\${^*}$ the majority of the earliest $\sim100$ observations were averaged to 2\,s and 24.4\,kHz due to the large data rates.
\end{table}

\begin{figure*}   \centering
   \includegraphics[height=5.35cm]{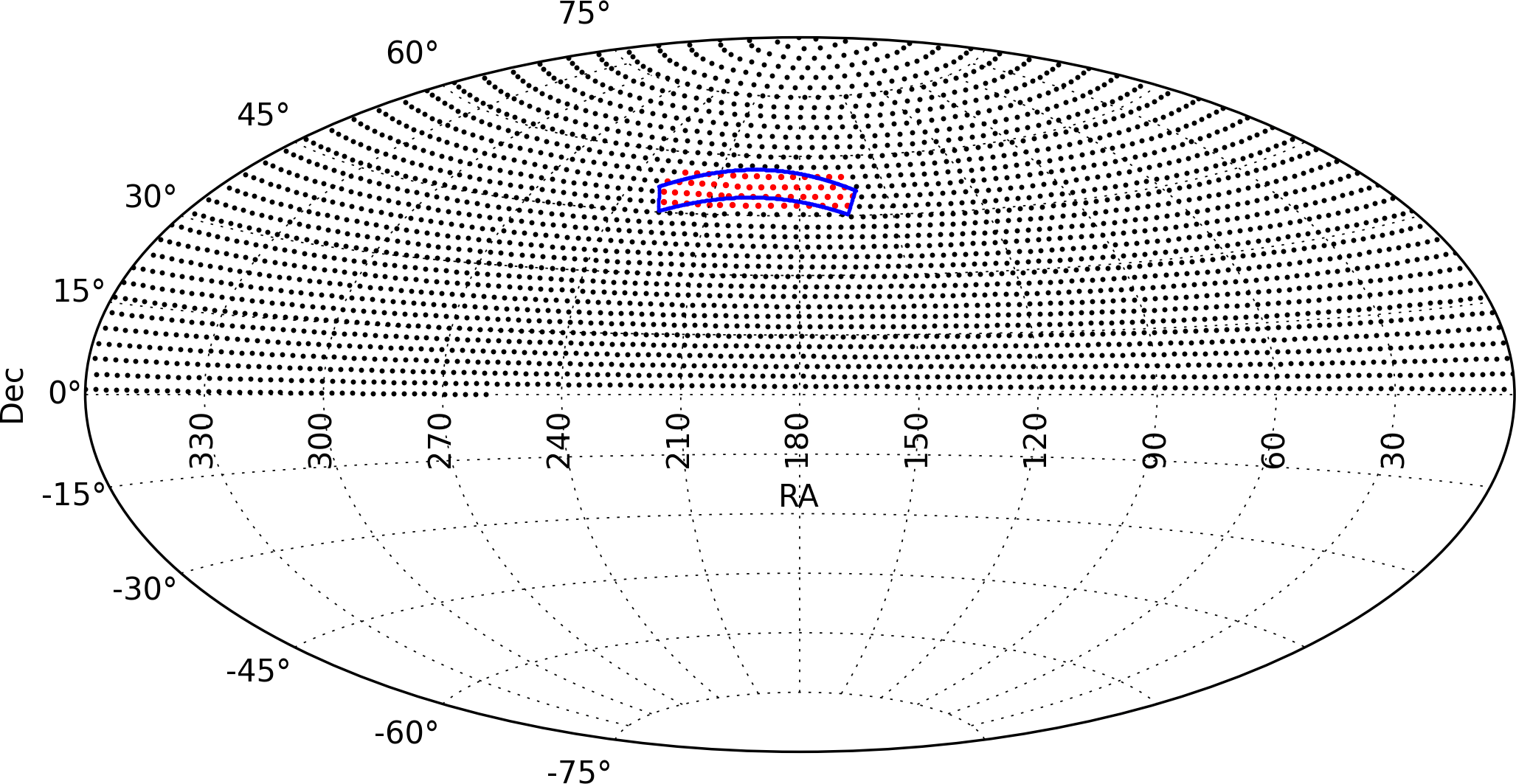}
   \includegraphics[height=5.35cm]{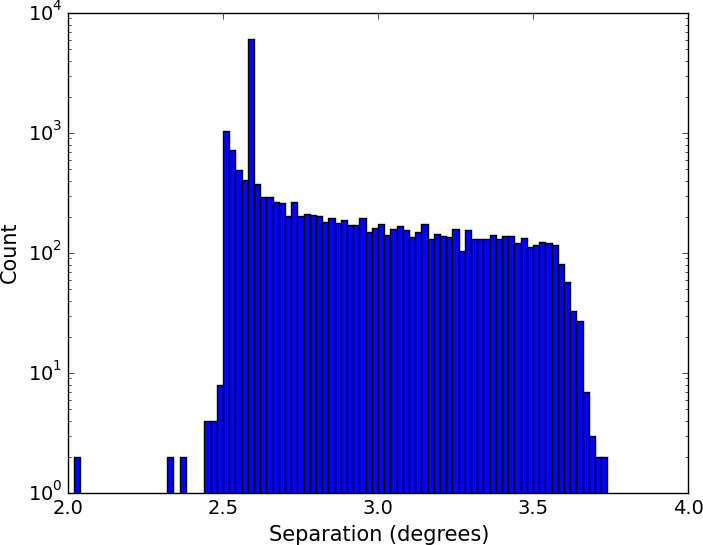}
   \caption{The left panel shows the LoTSS pointing grid which follows a spherical spiral structure. The region highlighted in blue is the HETDEX Spring Field. The red points show the LOFAR pointings that are presented in this publication and the black points show the rest of the survey grid. The right panel shows a histogram of the separation of the six nearest neighbours to each of the 3170 pointings in the survey grid excluding the edge pointings close to declination zero. A log scale is used on the y-axis in order to clearly display the full variation of pointing separations. {The mean separation of pointings is 2.80$^\circ$ but the distribution is highly peaked around the median separation of 2.58${^\circ}$. In total, 65\% of pointings have all six nearest neighbours within 2.80${^\circ}$ and 98\% have at least four neighbouring pointings within 2.80${^\circ}$.} Note that the panel on the right was created from a grid with a complete spherical spiral structure and ignores the 42 test pointings that were conducted with a slightly different tiling strategy.}
   \label{fig:Tier-1-HBA-pointings}
\end{figure*}

\section{Observation status}
\label{sec:obsstatus}

The LoTSS was initiated on 2014 May 23 in the region of the HETDEX Spring Field and in this publication we present preliminary images of the surveyed region between right ascension 10h45m00s to 15h30m00s and declination 45$^\circ$00$\arcmin$00$\arcsec$ to 57$^\circ$00$\arcmin$00$\arcsec$ (see Figure \ref{fig:Tier-1-HBA-pointings}) that encompass the HETDEX Spring Field. Our observations of this field comprise 63 pointings that were observed between the start of the survey and 2015 October 15. Each pointing was observed for approximately 8\,hrs and a calibrator (3C 196 or 3C 295) was observed before and after the observation of the target. 

The 63 LoTSS pointings within the region of the HETDEX Spring Field are only 2\% of the total survey. However, by 2016 November we will have gathered data for 350 LoTSS pointings whose coverage spans far beyond the HETDEX region. Our top priority is to complete the survey above declination $>25^\circ$, where the sensitivity of LOFAR is highest: the existing observations correspond to 20\% of this region. At the current rate of observations we expect to complete at least this region with the next 5 years.

\section{Data reduction}
\label{sec:datareduction}

The reduction of the LoTSS data is challenging due to: the large data size; the desire to reduce the data to approximately match the rate at which new observations are performed; the need for almost complete automation; and the complexities involved in calibrating the direction-dependent ionospheric effects and beam model errors. Here we present a preliminary reduction of LoTSS data that was performed with a completely automated {\it{direction-independent}} calibration and imaging pipeline that we describe in detail in the following subsections. This calibration allows us to create 25$\arcsec$ resolution images with a noise level that is typically in the range from 200 to 500\,$\mu$Jy/beam away from bright sources. However, we emphasise that in the longer term, we will complete a full {\it{direction-dependent}} calibration of these data that will enable us to reach the thermal noise of approximately 100\,$\mu$Jy/beam at a resolution of 5$\arcsec$. One such procedure to produce the desired high quality images from similar datasets was recently outlined by \cite{vanWeeren_2015a} and \cite{Williams_2015}. At present, this procedure requires too much user interaction and computational time to be routinely run on the LoTSS datasets but good progress is being made to reduce these requirements.

\subsection{Calibration}
\label{sec:calibration}
The direction-independent calibration procedure we have adopted is similar to that applied in preparation for the direction-dependent facet calibration scheme developed by \cite{vanWeeren_2015a} and \cite{Williams_2015}. The difference is that we apply the standard LOFAR station beam model during the imaging using AWimager (\citealt{Tasse_2013}). For completeness the direction-independent calibration strategy is outlined below.

The data for the target ($\approx8$\,hrs) and the calibrator ($2\times10$\,mins) were recorded with 1\,second sampling  and 64 channels per 0.195\,MHz subband. These data were flagged for interference by the observatory using the \textsc{AOFLAGGER} (\citealt{Offringa_2012}) before being averaged. Only the averaged data products, which have sizes between 3\,TB and 16\,TB per pointing (depending on the averaging), were stored in the LOFAR archive. 

Prior to calibration, the data were downloaded from the {LOFAR Long Term Archive} to local computing facilities at a speed of about 30MB/s. At this speed, the retrieval of a 3\,TB dataset took $\approx$1day and a 16\,TB dataset took $\approx$1\,week. After the data were retrieved from the archive, we averaged the calibrator data to 4 channels per 0.195\,MHz subband and 4 seconds, flagged again for interference (which is identified by \textsc{AOFLAGGER} on the XY and YX polarisations) and removed the international stations from the measurement set if they were included in the observation. Each subband of the calibrator data was then calibrated using the \textsc{BLACKBOARD SELFCAL (BBS)} software (\citealt{Pandey_2009}) to obtain XX and YY solutions for each time slot and frequency channel, taking into account differential Faraday rotation. In these data the only calibrators observed were 3C 295 and 3C 196, and these were used to calibrate 4 and 59 pointings respectively. The model used for the calibration of 3C 295 uses the flux density scale provided by \cite{Scaife_2012} with the flux density split equally between two point source components separated by 4$\arcsec$. The model used for the calibration of 3C 196 is also consistent with the flux density scale described in \cite{Scaife_2012}, consisting of a compact ($<6\arcsec$ maximum separation) group of four narrow gaussian sources (with major axis less than 3$\arcsec$) each with a spectral index and curvature term (V. N. Pandey, private communication). 

After each subband of the calibrator data has been calibrated, the calibration tables for all 244 subbands are combined into a single table for all 48\,MHz of available bandwidth. Using the full-bandwidth calibration table, we smooth the XX and YY amplitude solutions in time and frequency to provide a frequency-dependent but time-independent amplitude solution for each station. These solutions are fairly stable with variations of $\approx$10\% over the 18 months that these observations were taken. {The exact cause of these variations is uncertain but likely includes the stability of the instrument, the elevation of the calibrator, the observing conditions, and the accuracy of the calibrator sky models}. In Figure \ref{fig:HETDEX-calibration-amplitude} we show example amplitude solutions for all observations within the HETDEX region for a representative sample of four LOFAR stations, including two core stations and two remote stations. 

The full-bandwidth calibration solutions span a sufficiently wide frequency range to allow us to separate the effects of the LOFAR clocks that timestamp the data prior to correlation (each remote station has its own clock and the core stations operate using a single clock) from those of the  Total Electron Content (TEC) difference following the scheme described in \cite{vanWeeren_2015a}. These effects can be separated as the clock difference between the stations causes a phase change that is proportional to $\nu$, whereas the difference in TEC between the lines of sight of the two stations causes a phase change that is proportional to $\nu^{-1}$. Example clock solutions are shown in Figure \ref{fig:HETDEX-calibration-clock}. This shows that the clock values for the core stations are around 0\,ns (this is by definition as the plots show the difference between the clocks of each station and the core station CS001HBA0) but the clock values for the remote stations can be $\approx 100$\,ns. Whilst we find that the clock solutions are generally quite stable, we do see small variations between observations. For example, for the remote stations, we find that there are two discrete groups of clock values (see Figure \ref{fig:HETDEX-calibration-clock}) and that these correspond to Cycle 2 and Cycle 3 observations (where each Cycle corresponds to 6 months of observations) between which the delay calibration was refined by the observatory. Furthermore, there are still variations within the derived clock values for observations within the same Cycle. This is expected because the remote stations have their own clocks, synchronised with a Global Positioning System (GPS) signal, and are known to drift by within $\sim$15\,ns time-scales during an observation as was demonstrated by \cite{vanWeeren_2015a}.

Similarly to the calibrator field, the target field is averaged to 4 channels per 0.195\,MHz subband and 4 seconds, flagged again for interference which is identified on the XY and YX polarisations and the international baselines are removed from the measurement sets. From almost all our HETDEX observations, the station CS013 is also flagged because until October 2015 the HBA dipoles of this station were rotated at 45$^\circ$ with respect to the other stations. The time independent clock values and amplitude solutions that were derived from the calibrator observations are then applied to the target data. The transfer of the clock and amplitude values is done at this step, prior to the full averaging of the target data, to reduce decorrelation that the clock offsets may cause on the longest baselines. The target data are then averaged by a further factor of 2 in both time and frequency to give a final frequency and time resolution of 2\,channels per subband and 8\,seconds. In Section \ref{sec:archived-data} we highlighted the need for less averaging (4\,s and 4 channels per subband) when imaging at 5$\arcsec$ resolution (see also \citealt{Williams_2015}) but in this preliminary data release our imaging is at a much lower resolution of 25$\arcsec$ and averaging to 2 channels per subband  and 8\,seconds causes minimal time or bandwidth smearing in the field of view. In our images of each pointing, the measured peak brightness 2.5$^\circ$ from the pointing centre should be 98\% of their expected value. However, we note that our pointings are mosaiced to produce the final images (see Section \ref{sec:mosaicing}). Sources in our mosaiced images will all have a reduced peak brightness due to smearing and the reduction will depend upon the position of the source with respect to each of the  pointing centres as well the weighting of each pointing in the mosaiced image (see e.g. \citealt{Prandoni_2000}). We have calculated that for sources detected in the central part of our mosaiced region (in pointings with six surrounding pointings; see Section \ref{sec:mosaicing}) the peak brightness loss will be less than 2\%, whilst for sources close to the outer edge of the mosaiced region the peak brightness loss remains below 4\%.

Due to the wide-field of view and the non-negligible sidelobes of the LOFAR HBA beam it is common that sources in distant sidelobes contribute significant artefacts across the main lobe of the beam. The primary cause of such emission is due to the very bright sources Cygnus A, Cassiopeia A, Virgo A, Taurus A and Hercules A. The contamination from these sources is assessed for each pointing by using models of the sources and the LOFAR HBA beam to simulate the response of each of them throughout the observations. These sources are all further than 35$^\circ$ from the pointings in the HETDEX Spring Field region, and due to this large separation we are able to efficiently minimise the contamination from them by simply flagging baselines and time periods where their simulated signal exceeds the observatory-recommended threshold of 5\,Jy. 

{After the bright contaminating sources were removed}, the target field data was concatenated into groups of 12 subbands (2.3\,MHz) and flagged for interference again with AOFLAGGER with a strategy that uses the XY and YX polarisations to remove low level interference that was not previously identified. The target data were then phase calibrated with a calibration time interval of 32 seconds against a sky model generated from the VLA Low-Frequency Sky Survey (VLSSr; \citealt{Lane_2012}), Westerbork Northern Sky Survey (WENSS; \citealt{Rengelink_1997}) and the NRAO/VLA Sky Survey (NVSS; \citealt{Condon_1998}) -- see The LOFAR Imaging Cookbook\footnote{https://www.astron.nl/radio-observatory/lofar/lofar-imaging-cookbook} or \cite{Scheers_2011} for  details. All VLSS sources within five degrees of the pointing centre with a flux density greater than 1\,Jy are included in the phase calibration catalogue and these sources are matched with WENSS and NVSS sources to include the spectral properties of the sources in the phase calibration catalogue. We note that imperfections in the sky model will result in calibration errors and efforts are ongoing to reduce these imperfections by utilising models derived from other surveys such as TGSS (\citealt{Intema_2016} and MSSS (\citealt{Heald_2015}). However, even with the sky model we presently use we often find that direction dependent effects, rather than sky model imperfections, are the primary limitation of the image quality (see Section \ref{sec:imaging}).

The control parsets and scripts that have been developed to perform the entire calibration procedure that is described above are executed by the pipeline framework that is now part of the LOFAR software package. Using this pipeline framework makes it simple to efficiently run our completely automated reduction on multiple computers. The pipeline framework handles data tracking, parallel execution, and checks each step is properly completed, which allows for jobs to be resumed. During the pipeline run, various diagnostic plots are produced to assess the quality of the data. For the calibrator observations we ensure that the values derived for the amplitude and clock corrections are good. We also examine the phase solutions from the target to quickly identify observations that suffer from poor ionospheric conditions. After the data are retrieved from the archive, approximately 3 days are required to execute this calibration pipeline on 24 threads of one of our compute nodes. Each of our compute nodes have 512 GB RAM and contain four Intel Xeon E5-4620 v2 processors which have eight cores each (16 threads) and run at 2.6\,GHz.

The final step is to remove time periods during which the ionospheric conditions are poor. We identify such conditions by locating time periods that have rapid large variations in phase. The phase calibration of the target provides a solution for each station every 32 seconds and, generally, when a nearby station is used for a phase reference, these solutions change smoothly as a function of time. Hence, the difference between these solutions and the same solutions smoothed along the time axis (using a median filter with a window size of 5 samples) is close to 0 radians for short baselines. Therefore, for each station we use the closest station as a reference for the phase solutions and identify periods of rapidly varying phases which are those where the difference between the raw solutions and the smoothed solutions are significant (we set a threshold of 0.29\,radians for a 12 subband dataset). If, for multiple stations (we use a threshold of 5 stations), we identify the same time period as having a rapidly varying phase the ionospheric conditions are classified as poor and the data are flagged for all stations. We note that that this technique works well if we only use the phase solutions from the core LOFAR stations, where the maximum distance to the nearest station that is used for a phase reference is 1675\,m (at this distance the phase solutions do not vary rapidly in normal observing conditions). As the remote stations are isolated, with no other stations nearby, there are often very rapid variations in the phase solutions when the nearest station is used as a phase reference (see e.g. \citealt{vanWeeren_2015a}) and poor ionospheric conditions can be more difficult to identify. This procedure to flag time periods with poor ionospheric conditions is demonstrated in Figure \ref{fig:HETDEX-P10-phases}.

\begin{figure}   \centering
   \includegraphics[width=\linewidth]{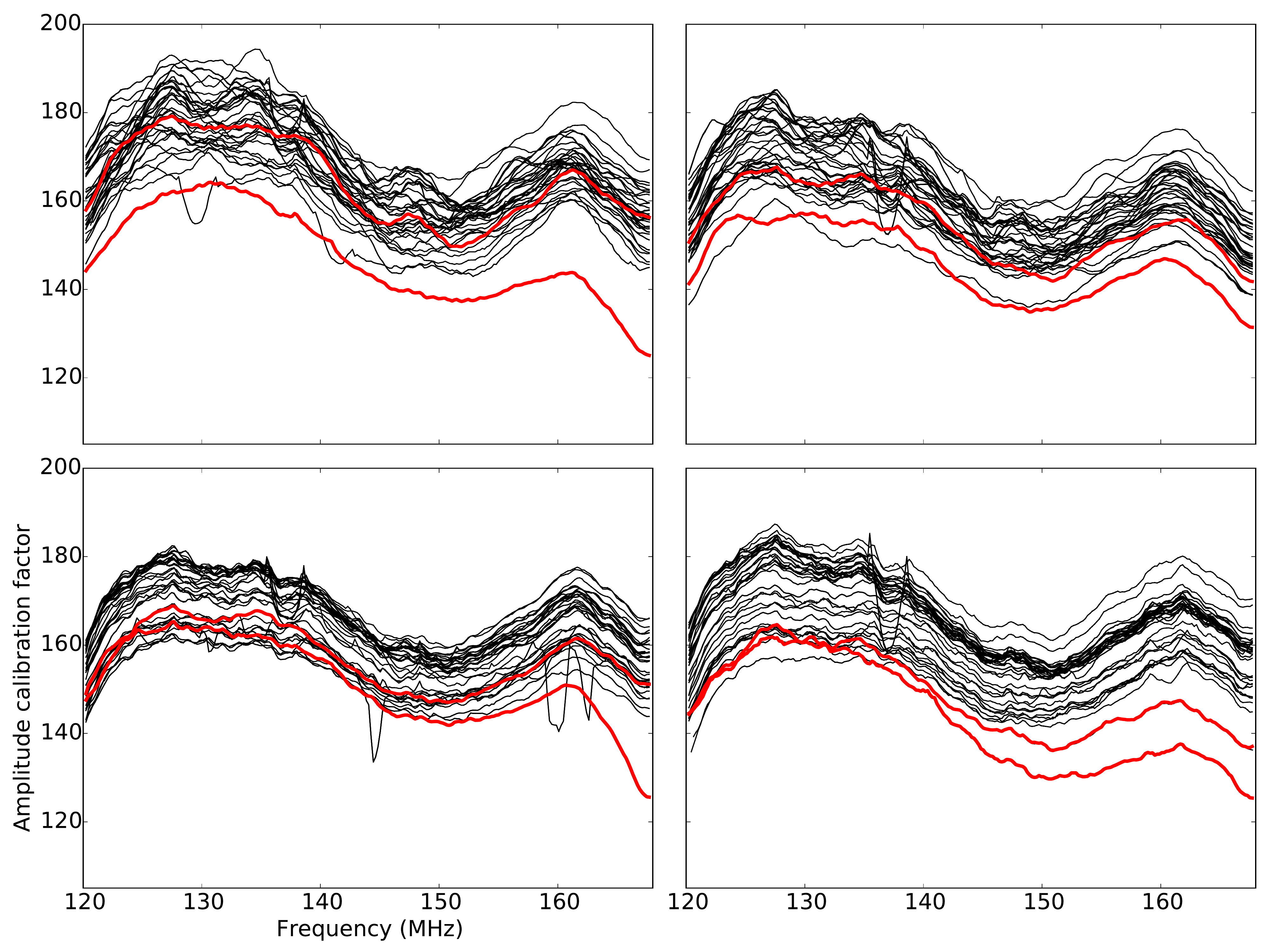} 
    \caption{Amplitude calibration solutions as a function of frequency for the calibrator observations that were used to convert correlator units  to Jy for the observations in the HETDEX Spring Field region. The lines show the amplitude solutions for different calibrator observations. The red lines are the solutions when 3C 295 was used as the calibrator and the black lines are when 3C 196 was used. The panels show the amplitude calibration solutions for two core stations (CS) and two remote stations (RS), from the top left these are: CS003HBA0, CS026HBA0, RS305HBA, RS509HBA. Several calibrator observations show small frequency ranges where bad data results in sharp changes in the amplitude solutions.}
   \label{fig:HETDEX-calibration-amplitude}
\end{figure}

\begin{figure}   
\centering
   \includegraphics[width=\linewidth]{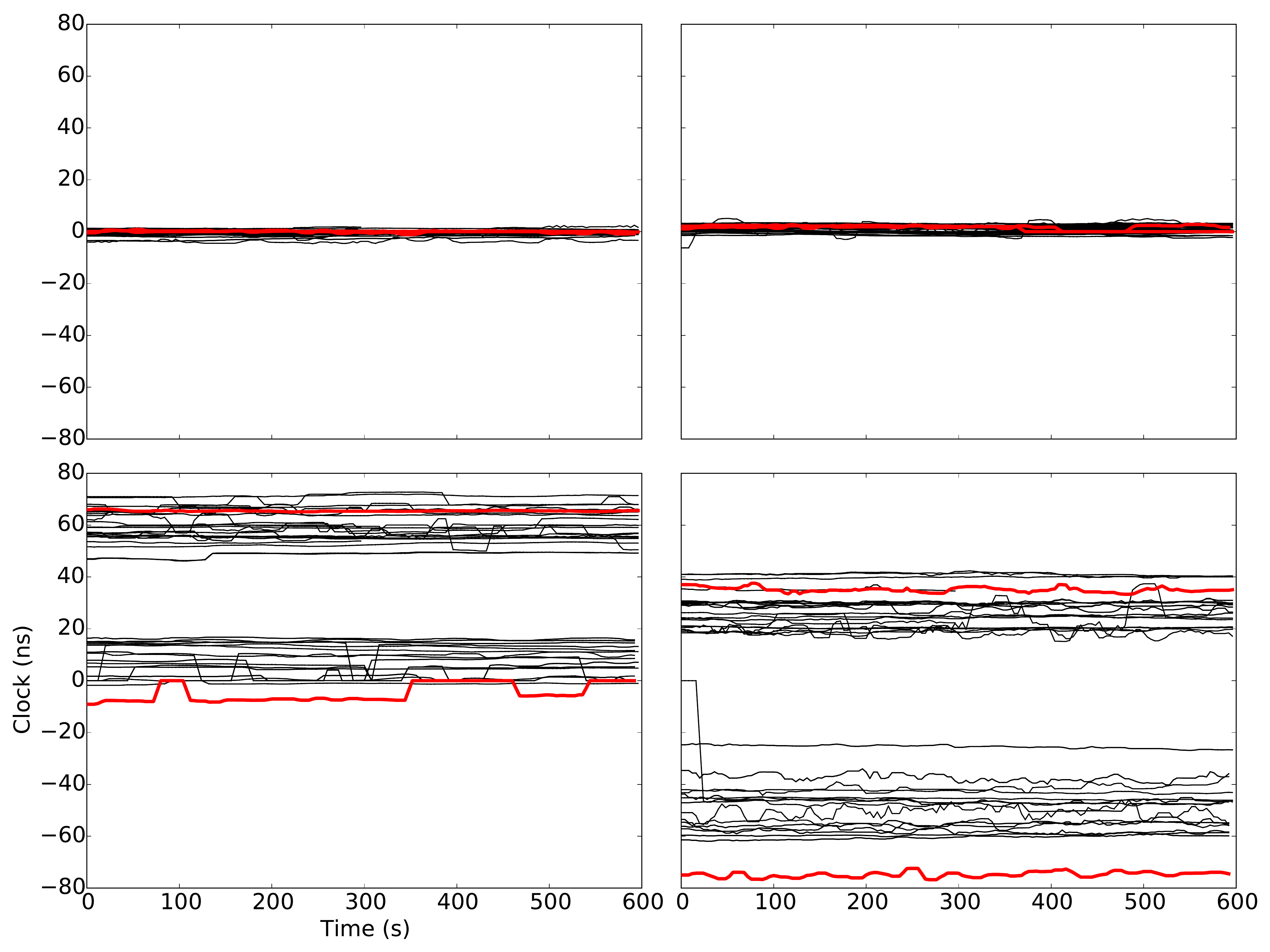} 
    \caption{Clock offsets as a function of time for the calibrator observations that were used to calibrate observations in the HETDEX Spring Field region. The lines show the clock offsets for different calibration observations. The red lines are the clock solutions when 3C 295 was used as the calibrator and the black lines are when 3C 196 was used. The panels show the clock offsets for two core stations (CS) and two remote stations (RS),  from the top left these are: CS003HBA0, CS026HBA0, RS305HBA, RS509HBA. There are several discontinuities in the derived clock values which are due to difficulties in converging on the precise clock solution (see \citealt{vanWeeren_2015a}), but only the median clock solutions are applied for calibration of the target field.} 
   \label{fig:HETDEX-calibration-clock}
\end{figure}

\begin{figure}   \centering
   \includegraphics[width=\linewidth]{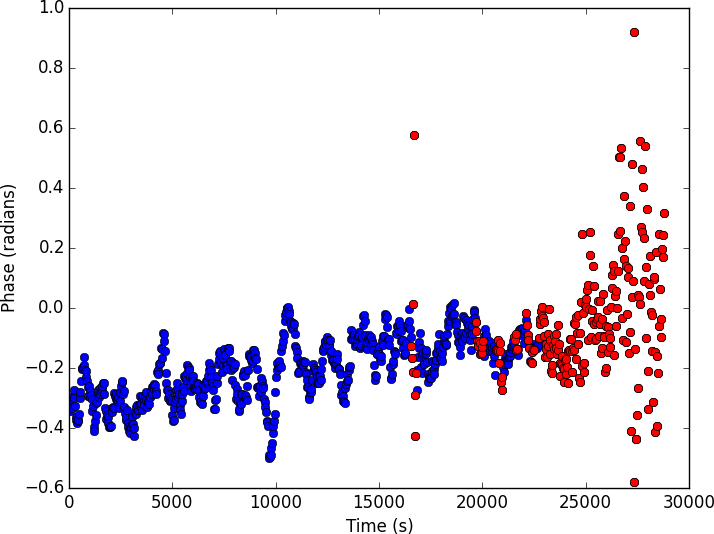}
    \caption{The phase solutions for station CS401HBA1 using station CS032HBA1 as a phase reference for a LoTSS dataset are shown in blue (CS032HBA1 is the closest station to CS401HBA1 at a distance of 584\,m). The red points show the time periods where the phase solutions indicate poor ionospheric conditions (see Section \ref{sec:calibration}) and these time periods are subsequently flagged.}
   \label{fig:HETDEX-P10-phases}
\end{figure}

\subsection{Imaging}
\label{sec:imaging}

We have somewhat mitigated direction-dependent effects by not utilising the full resolution of the Dutch stations of LOFAR ($\approx5\arcsec$) and only using baselines shorter than 12\,k$\lambda$ (corresponding to $\approx25\arcsec$ resolution) when imaging. However, wide-field imaging of these direction-independent calibrated LOFAR datasets is still difficult due to the low dynamic range of the images and the large number of bright sources. The high sidelobes of the LOFAR synthesised beam {($\sim12\%$ when imaging our data using the \citealt{Briggs_1995} weighting scheme and a robust parameter of $-0.5$)} can further hinder this procedure.  Furthermore, we use the AWimager (\citealt{Tasse_2013}) to apply the {time dependent} LOFAR station beam model in the imaging procedure to output both primary-beam corrected and uncorrected images, but with this imager we were unable to image all 48\,MHz of bandwidth with a single wide-band \textsc{clean} due to the large amount of data ($\sim$250\,GB of data per pointing), and at the time of writing, multi-frequency deconvolution was not supported. Such a wide-band deconvolution would be preferable as the synthesised beam sidelobes would decrease and it would be easier to identify and \textsc{clean} faint sources. Instead, we image 36 subbands together and create seven images with frequencies approximately evenly spaced across the 120-168\,MHz bandwidth (the highest frequency of these seven images consists of $\approx$28 subbands rather than 36). To efficiently \textsc{clean} the faint sources in the presence of large artefacts around bright sources we perform an automated multi-threshold \textsc{clean} where we progressively remove  \textsc{clean} boxes around bright sources to allow for the faint sources to be properly deconvolved, as described in detail below. {Throughout this imaging procedure we weight the visibilities with a robust parameter equal to $-0.5$ and image an area of $6.5^\circ \times6.5^\circ$ to ensure that bright sources far down the beam are deconvolved.}
 
We initially \textsc{clean} our {Stokes I} image to a threshold of 20\,mJy/beam without using a \textsc{clean} mask. The PyBDSM source finding software (\citealt{Mohan_2015}) is then run on the resulting deconvolved apparent brightness image that has an approximately uniform noise across the imaged area, but due to the limited dynamic range there are regions of increased noise around bright sources. This is used to create a \textsc{clean} mask that contains islands that tightly encompass all sources detected in the image but not artefacts around bright sources or source sidelobes, and a noise map that accurately describes the local noise at each position in the image. To approximately \textsc{clean} the image to the local noise at each position, we first \textsc{clean} the entire image using the \textsc{clean} mask and a threshold that is either the largest {noise} measurement on the {PyBDSM generated noise map} or 20\,mJy/beam (whichever is less). After this deeper \textsc{clean}ing of the entire field, the brightest sources are essentially fully deconvolved because the local noise is higher in those regions, but the fainter sources are not. Therefore, all pixels where the {noise map} value exceeds a given threshold are removed from the \textsc{clean} mask and the deconvolution is continued to a lower noise level. To properly \textsc{clean} the faintest sources to the local noise we repeat this procedure three times. This progressively removes the bright sources where the local noise is higher from the \textsc{clean} mask and lowers the \textsc{clean} threshold until {only the faintest sources are left in the \textsc{clean} mask and the threshold} reaches approximately the median value of the {noise map}. In a few cases, where the images contained very bright sources, we manually tweaked the imaging \textsc{clean} thresholds to improve the deconvolution. 

To create full bandwidth images, the seven different images across the band were stacked in the image plane. To do this, for each pointing, the images are convolved with a Gaussian intensity distribution to give the seven different images across the band the same resolution. The seven images are then stacked together by taking a weighted average of the images where the weight is 1/$\sigma^2$ and the noise, $\sigma$, is measured from the image by fitting a Gaussian probability distribution to pixel values from the non primary beam corrected image and discarding outlying values.  Due to the varying amounts of data that are flagged for different pointings, as well as the occasional subbands missing due to telescope errors, the individual pointings consist of varying proportions of different frequency components. Therefore the average weighted frequency of the seven stacked images is naturally slightly different for each pointing, with the average being 149\,MHz and the standard deviation 1.5\,MHz. Whilst the weighting of the image stacking could be adjusted to give the same weighted average frequency, this would still not ensure that all images have precisely the same frequency coverage. 

It is desirable to provide images with a uniform resolution. However, the missing subbands, the data flagged, the observation duration, and the target position will all result in variations in the synthesized beam between observations. We find that our images typically have a synthesised beam major axis FWHM of approximately 20.2$\arcsec$ with a range from 17.8$\arcsec$ to 24.8$\arcsec$, apart from two outlier fields, P2 and P8, which have synthesised beams that exceed 30$\arcsec$. The large synthesised beams are because over 80\% of these datasets (which were observed simultaneously) were flagged due to poor ionospheric conditions that were identified by the flagging procedure outlined in Section \ref{sec:calibration}. Therefore, we exclude these two fields from further analysis. To make the images uniform in resolution we convolve the remaining 61 images with a Gaussian of appropriate size to make the beam of each image $25\times25\arcsec$.

We note that these LOFAR images could be used to obtain a model of the sky that is higher resolution and more sensitive than that used in the initial phase calibration, and that this model could be used to self calibrate the LOFAR datasets. However, this procedure was not followed because self-calibration is time consuming and, while there is a dependance on the quality of the initial sky model in the target region, in most cases it was not found to significantly improve the image quality when imaging at 25$\arcsec$ resolution. This lack of a significant improvement in image quality is probably due to direction-dependent effects, rather than imperfections in the sky models that are used for the direction-independent phase calibration, dominating the calibration errors and limiting the image fidelity. In addition, the images could have been used to identify the sources that produced the largest artefacts (such as 3C 295), that could then be removed by constructing good models for the sources and using the peeling technique  (see \cite{Mahony_2016} for an example of peeling a bright source in LOFAR direction-dependent calibrated images). This operation was not performed due to the large number of sources that would require peeling and the computational expense associated with this. 

\section{Image quality}
\label{sec:imagequality}

The 25$\arcsec$ resolution images produced from our datasets form \newpage \noindent the most sensitive wide-area low-frequency survey yet produced (see Figure \ref{fig:comp_sensitivity}). The quality of images varies significantly between pointings due to the presence of bright sources in the field and the quality of the input sky model, but it is predominantly dictated by the position- and time-varying ionospheric conditions that cannot be corrected by a direction-independent calibration. This prevents accurate high-resolution imaging, as the ionosphere introduces phase errors which cause position changes that are non-negligible in size compared to the synthesised beam. Even though we have only used baselines shorter {than 12\,k$\lambda$ when} imaging, the uncorrected ionospheric phase errors  cause a noticeable blurring of sources, which reduces their peak brightness, alters their position and increases the image noise.  Furthermore, the quality of all {our images} is significantly hindered by imperfections in the LOFAR beam model which result in large direction-dependent amplitude (i.e. flux density and spectral index) variations as a function of time. The magnitude of all the quality variations amongst images will be reduced substantially once direction-dependent calibration is fully implemented. However, it is likely that poor ionospheric conditions will mean that a large number of directions will be required to properly calibrate an affected dataset. It may even be the case that, for some pointings, the ionosphere is so spatially variable that there is insufficient flux density within each isoplanatic patch to allow the calibration of all directions. Alternatively, it could be that the number of directions becomes so large that the number of degrees of freedom required for calibration approaches or exceeds the number of independent measurements of visibilities. Pointings where the ionospheric conditions prohibit a full direction-dependent calibration must be re-observed.

In the following subsections, we use LOFAR source catalogues for each pointing (created using PyBDSM) to first identify observations conducted in poor ionospheric conditions and exclude these from future analyses before we assess the quality of each of our remaining images by measuring the astrometry of compact objects, the flux density accuracy, and the sensitivity.

\subsection{Identifying poor ionospheric conditions}
\label{sec:identifying-bad-ionosphere}

An effective proxy for the ionospheric-induced blurring of sources during LOFAR observations is the ratio of the measured integrated flux density to the peak brightness. This is because the blurring substantially reduces the peak brightness while (except in very poor conditions) the integrated flux density is nearly preserved. Therefore, quantifying this ratio for each pointing allows us to identify and remove the observations that were conducted in the poorest ionospheric conditions. This procedure is simplified if just compact and isolated sources are used: for compact sources we expect the peak brightness and integrated flux density to be comparable and only selecting isolated sources reduces the probability of mismatched sources or artefacts in the catalogue. To create such a sample of sources for each pointing, we match the LOFAR catalogue with the FIRST catalogue which is used because it has a high resolution ($\approx5\arcsec$) and helps identify  compact sources. The cross matching is performed by simply matching all LOFAR and FIRST sources that are within 10$\arcsec$. Entries are removed from this cross matched catalogue if they are: within 30$\arcsec$ of another LOFAR detected source;  further than 2$^\circ$ from the LOFAR pointing centre; have multiple matches; or have sizes greater than 10$\arcsec$ in the FIRST catalogue or greater than 30$\arcsec$ in the LOFAR image.

The integrated LOFAR flux density divided by the peak LOFAR brightness for all objects in our cross matched catalogues is shown in Figure \ref{fig:lofar-flux-ratios}. We find that the typical median value of this ratio of compact sources for a pointing is 1.2 but for the 61 pointings we are analysing it varies from 1.1 to 2.0. There are seven pointings (P6, P164+55, P21, P225+47, P206+50, P221+47 and P33) that we identified as having particularly high integrated flux density to peak brightness ratios (with a median exceeding 1.35) indicating substantial ionospheric blurring. These pointings are excluded from the remainder of this study, which leaves 54 pointings for further analysis.

\begin{figure}   
\centering
   \includegraphics[width=\linewidth]{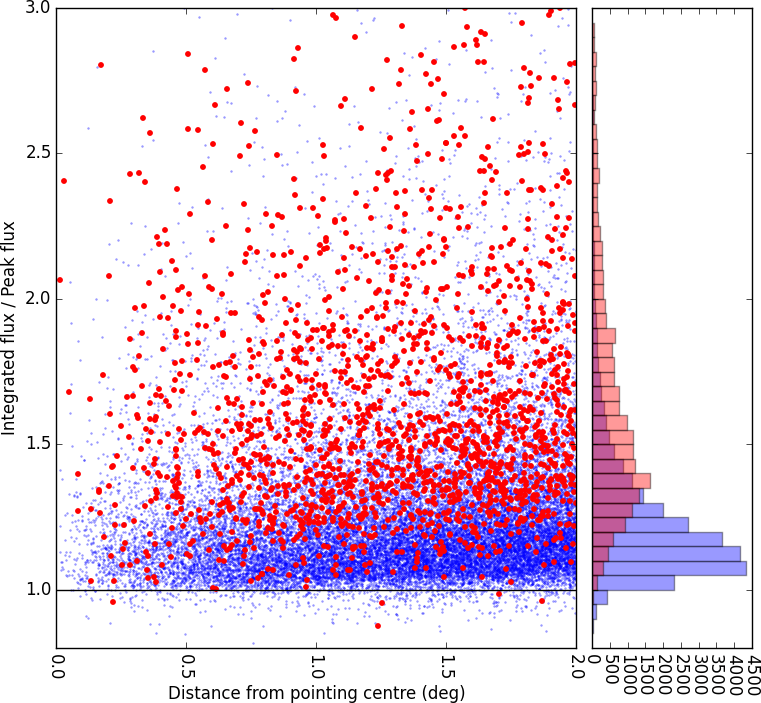}
    \caption{The ratio of the integrated flux density to peak brightness for compact sources in all 54 LOFAR pointings. The seven pointings identified as having particularly poor ionospheric conditions are shown in red and the remaining 54 pointings are shown in blue. The histogram of the red points has been multiplied by a factor of 10 for display purposes.}
   \label{fig:lofar-flux-ratios}
\end{figure}

\subsection{Astrometric uncertainties}
\label{sec:astrometry}

The astrometry of our images is set by our phase calibration, in which we use a model created from the VLSS, WENSS and NVSS surveys (see Section \ref{sec:calibration}). These surveys are at lower resolutions than ours and inaccuracies in the model will not be uncommon. For example, there will be double sources that are unresolved in the lower-resolution model but resolved in our higher-resolution datasets and there will be complex extended emission that is poorly characterised in the model. These imperfections in the phase calibration catalogue will result in a systematic error in the position of our sources and this will vary between pointings. Furthermore the final astrometric {accuracy of our images} can also be affected by inaccuracies in the beam model and the ionospheric conditions during the observation. In our images we have not attempted to correct the astrometry for  direction dependent calibration effects but we are able to correct  systematic position offsets.

To examine the astrometry of our images and correct the systematic astrometric offset for each pointing, we again cross match catalogues of sources from each LOFAR pointing with the FIRST catalogue. The FIRST catalogue was used as it has systematic position errors of less than 0.1$\arcsec$ from the absolute radio reference frame, which was derived from high resolution calibrator observations (\citealt{White_1997}). The cross-matching is performed using exactly the same procedure as was described in Section \ref{sec:identifying-bad-ionosphere}. Thus, the final cross matched catalogue contains only compact and isolated sources and this alleviates the issue of possible source brightness distribution changes between the 150\,MHz LOFAR  and 1.4\,GHz FIRST measurements.  

The final cross-matched catalogue was used to correct the systematic position offset within each LOFAR pointing. This was done by using the median right ascension and declination offsets ($\Delta$RA and $\Delta$Dec) to align the LOFAR source positions with those measured in FIRST. During this process we progressively filtered out sources with offsets more than three median absolute deviations (MAD) from the median offset until the median offset converged. The calculated offsets, which range from $-3$ to 6$\arcsec$ in RA and $-6$ to 3$\arcsec$ in DEC, were then applied by altering the headers of the LOFAR image files. 

After the correction of the systematic position offset, the LOFAR catalogues were remade and again cross matched with FIRST using the same criteria. It is apparent from this cross matching that the quality of the direction-independent calibration of the LOFAR datasets still varies significantly, which is indicated by variations in both the number of LOFAR sources matched with FIRST sources after filtering out all sources that are not compact and isolated, and the standard deviation of the position offsets. Whilst these variations (e.g. a high standard deviation of the cross-matched source offsets or a low number of cross-matched sources) could be used to further identify observations conducted during poor ionospheric conditions where direction-dependent position offsets are large, we do not use them here. The final astrometric accuracy of the images we have produced through our direction-independent calibration pipeline is displayed in Figure \ref{fig:astrometry-post-correction}. We find that the standard deviation of the offsets, without filtering outliers, is 1.65$\arcsec$ in RA and 1.70$\arcsec$ in declination which is less than 10\% of the synthesised beam size and smaller than the image pixels. By comparison, the TGSS alternative data release, which is at a similar resolution to our LOFAR images but has direction-dependent ionospheric corrections applied, has a standard deviation of 1.55$\arcsec$ in the offsets between their measured source positions and those recorded in a VLBA calibrator catalogue (see Figure 13 of \citealt{Intema_2016}). The LOFAR MSSS verification field, which is at a lower resolution of 108$\arcsec$ and without a correction for direction dependent effects, has slightly larger offsets of 2.92$\arcsec$ in RA and 2.45$\arcsec$ in DEC from the NVSS source positions (\citealt{Heald_2015}).
 
\begin{figure}   
\centering
   \includegraphics[width=\linewidth]{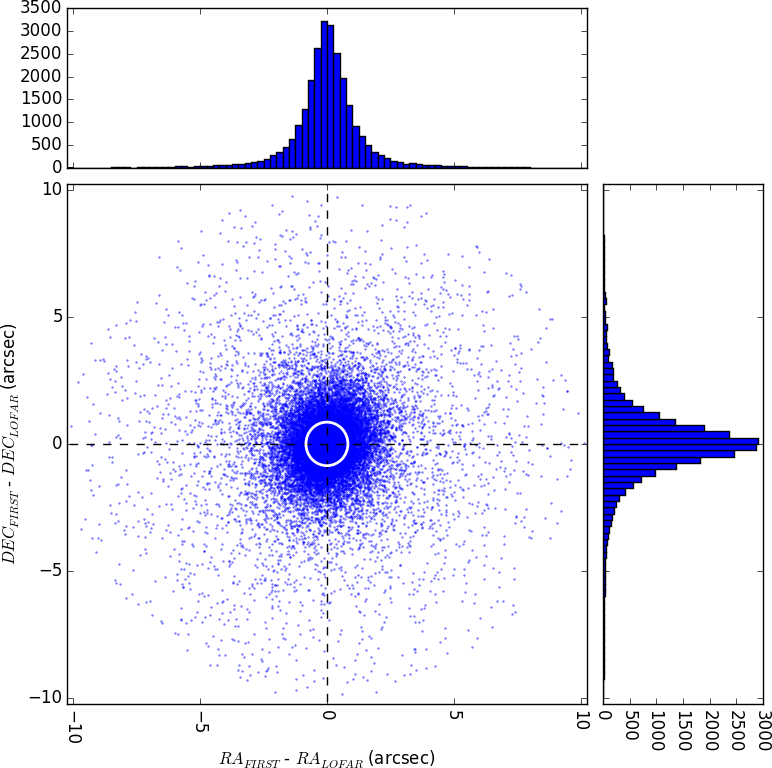}
    \caption{The residual RA and Dec offsets for LOFAR detected sources matched with their FIRST counterparts. The ellipse shows the standard deviation of the RA and Dec offsets (1.65$\arcsec$ and 1.70$\arcsec$ respectively) and is centred on the mean offset which is zero in both RA and Dec. The points show the RA and Dec offsets for each of the matched sources. The histograms show the number of sources at different RA and Dec offsets.}
   \label{fig:astrometry-post-correction}
\end{figure}

\subsection{Flux density uncertainties}
\label{sec:flux-comparison}
For amplitude calibration, we used models of 3C 196 and 3C 295 to calibrate 94\% and 6\% of the pointings respectively. The models for both calibrators are on the same flux density scale as the amplitude calibration models that were presented in \cite{Scaife_2012}. These models, even in the presence of the known imperfections in the LOFAR HBA beam model, should allow us to obtain flux density accuracies within 10\% (see e.g. \citealt{Heald_2015} and \citealt{Mahony_2016}). However, as we have not corrected for ionospheric phase errors, we expect that our flux measurements may be reduced due to a blurring of the sources, where the peak brightness will be affected significantly more than the integrated flux density as was quantified in Section \ref{sec:identifying-bad-ionosphere}.

To assess the overall errors on our 150\,MHz LOFAR integrated flux density and peak brightness measurements, we compared with the 7C and TGSS alternative data release measurements. After the astrometric correction of our images (see Section \ref{sec:astrometry}), we matched our LOFAR sources to these catalogues using the procedure that is outlined in Section \ref{sec:identifying-bad-ionosphere}, but as we are not matching with the FIRST catalogue here we did not filter our sources based on their size in that catalogue. Thus, similarly to the catalogues used for astrometric corrections, the cross-matched catalogues contain only compact and isolated sources. When matching with the low-resolution 7C catalogue we find a median ratio of the 7C integrated flux density (peak brightness) to the LOFAR integrated flux density (peak brightness) of 1.06 (1.26). Similarly, when matching with the approximately equal resolution TGSS catalogue we find the integrated flux density and peak brightness ratios to be 0.95 and 0.94 respectively (see Figure \ref{fig:flux-comparison}). 

To ensure that the inaccuracies in the LOFAR beam model do not result in significant systematic flux density errors we measured the variation in the TGSS to LOFAR integrated flux density ratio as a function of distance from the LOFAR pointing centre. Excluding the region within 0.15$^\circ$ of the pointing centre, which has a low number of cross matched sources, this was found to be small, with the measured median values ranging from 0.92 to 0.95 and all measurements out to 2.5$^\circ$ from the pointing centre agreeing within the errors (see Figure \ref{fig:flux-comparison}). Additionally, we found no clear trends in the integrated flux density ratio of sources as a function of right ascension or declination. These tests indicate that there are no obvious systematic flux density errors in our measurements of sources within the HETDEX region due to the LOFAR beam shape.

The variation in the LOFAR image quality is again reflected by the variation in the consistency between the LOFAR flux density measurements and those in other catalogues. For example, the TGSS to LOFAR median integrated flux density ratio for single pointings ranges from 0.82 to 1.26, although 95\% of pointings have values less than 1.1 and the MAD is only 0.05. Additionally, we compared the LOFAR integrated flux density measurements in the overlapping regions of neighbouring pointings, and found that the median ratio of the measurements in one pointing to those in neighbouring pointings varied from 0.85 to 1.12. While most of our images have flux density estimates within an uncertainly of 10\% some have larger uncertainties than this. Therefore, to reflect the variation in the LOFAR image quality, we put a conservative 20\% error on all flux estimates.

We note, that due to the tight mask used during the deconvolution of our LOFAR images we do not expect a large \textsc{clean} bias which would cause a systematic reduction in our flux measurements (e.g. \citealt{Becker_1995}) . However, we do expect that this bias varies significantly between pointings due to the uncorrected direction-dependent amplitude errors (see \citealt{Williams_2015}) and the large variation in the uncorrected direction dependent ionospheric effects. A detailed simulation to inject sources into our datasets and assess the level of \textsc{clean} bias should take into account these effects, but such a assessment is beyond the scope of this preliminary data release publication. A thorough evaluation of \textsc{clean} bias will be performed on the final direction dependent calibrated LoTSS data, although we note that an initial investigation was performed by \cite{Williams_2015} who found no significant \textsc{clean} bias.

\begin{figure*}   
\centering
\includegraphics[width=0.48\linewidth]{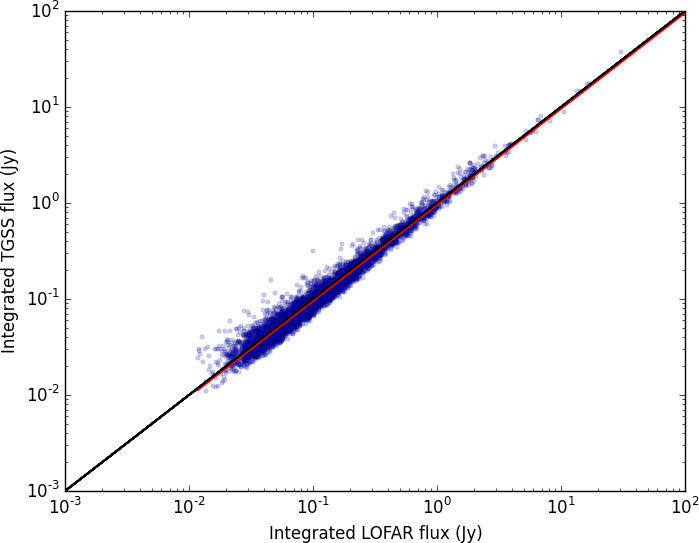}
\includegraphics[width=0.48\linewidth]{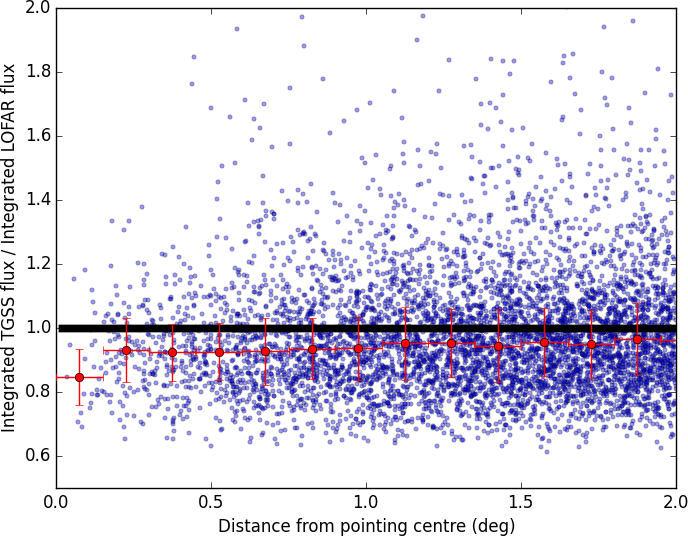}
 \caption{A comparison between the LOFAR integrated flux density measurements and the TGSS measurements with each cross-matched isolated compact source shown with a solid circle. On the left is a comparison of the integrated flux {densities. The} solid black lines show a 1:1 ratio of the integrated flux densities and the solid red line shows the median ratio between the integrated flux density measurements. {On the right is a comparison of the integrated flux density ratio as a function of distance from the LOFAR pointing centre. The red symbols indicate the median within bins of distance, with} the vertical error bars showing the median absolute deviation (MAD) value for each bin and the horizontal error bars giving the bin width. The median value of the TGSS integrated flux density divided by the LOFAR integrated flux density is 0.95.}
\label{fig:flux-comparison}
\end{figure*}

\subsection{Sensitivity}

Whilst we have removed the pointings with the worst ionospheric observing conditions (see Section \ref{sec:identifying-bad-ionosphere}), the sensitivity of our remaining images still varies significantly. This is due to imperfect calibration resulting in a limited dynamic range that leaves significant artefacts around bright sources whilst the uncorrected ionospheric phase errors scatter flux throughout the image.

To quantify the noise in our images, we fit a Gaussian to a histogram of image pixels after the array of pixels has been filtered to remove entries with values exceeding 10\,mJy/beam. A histogram of the measured noise values for the 54 pointings within the HETDEX field is displayed in Figure \ref{fig:noise-histogram}. We find that the median noise level is 380\,$\mu$Jy/beam and the range is from 270\,$\mu$Jy/beam to 960\,$\mu$Jy/beam where the pointings with the highest noise are around the very bright calibrator source 3C295 and dynamic range limitations result in a high noise value. We note that although the image fidelity in the direction-independent calibrated images is low the sensitivity we achieve in the best of these images is comparable to that obtained at 25$\arcsec$ using a direction-dependent calibration scheme such as facet calibration. The reason that images from direction-dependent calibrated data are often significantly more sensitive is that more baselines are used. Whilst we have removed baselines longer than 12\,k$\lambda$ because we are unable to reliably calibrate them with direction-independent calibration, the high resolution ($\approx 5 \arcsec$) direction-dependent calibrated images that reach a sensitivity of $\approx100\,\mu$Jy/beam use all the stations within the Netherlands. However, the worst of our images has a noise level that is approximately five times higher than the noise that would be expected from imaging the same baseline range but after direction-dependent calibration.

\begin{figure}   
\centering
   \includegraphics[width=\linewidth]{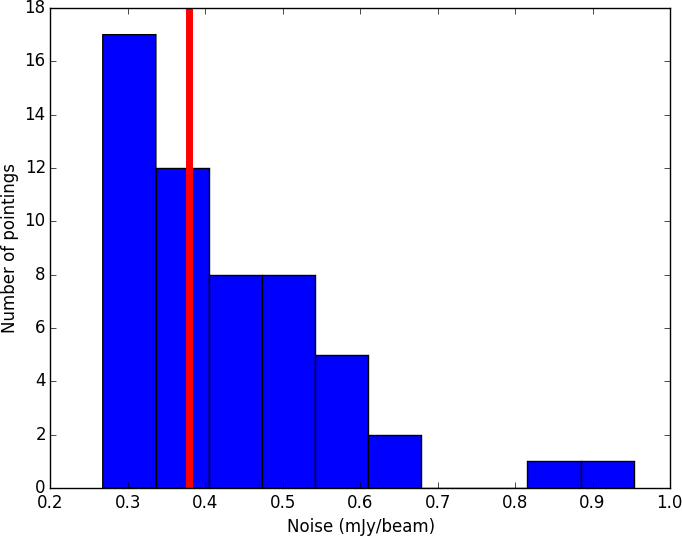}
    \caption{A histogram showing the noise estimates from the direction-independent calibrated LOFAR images. The median noise level (shown by the vertical red line) for the 54 HETDEX pointings is 380\,$\mu$Jy/beam and the noise levels range from 270\,$\mu$Jy/beam to 960\,$\mu$Jy/beam.}
   \label{fig:noise-histogram}
\end{figure}

\section{Mosaicing}
\label{sec:mosaicing}

As described in Section \ref{sec:survey-strategy}, neighbouring pointings in the LoTSS overlap at approximately the FWHM/$\sqrt{2}$ and therefore mosaicing the images from neighbouring pointings significantly improves the sensitivity compared to the images from single pointings. After the astrometry corrections have been applied we construct 54 mosaiced images, one centred on each of the 54 pointings. The map value at any point on the mosaiced images is derived from the pixels on each constituent primary beam corrected map, $m_{i}$, the primary beam value at each pixel $p_i$, and an estimate of the central noise level for each map, $\sigma_i$. The pixels from each constituent map are added together and weighted according to the noise, where the weight is given by $\big(\frac{p_i}{\sigma_i}\big)^2$. Two large mosaics, each showing  half of the HETDEX Spring Field region, are shown in Figures 16 and 17. We note that in these images a few of the nine excluded pointings are apparent at the south eastern edge of the mosaic, but in the central region of the mosaic the large overlap over pointings makes the raised noise level due to excluded pointings difficult to identify.

\begin{sidewaysfigure*}   
\centering
   \includegraphics[width=\linewidth]{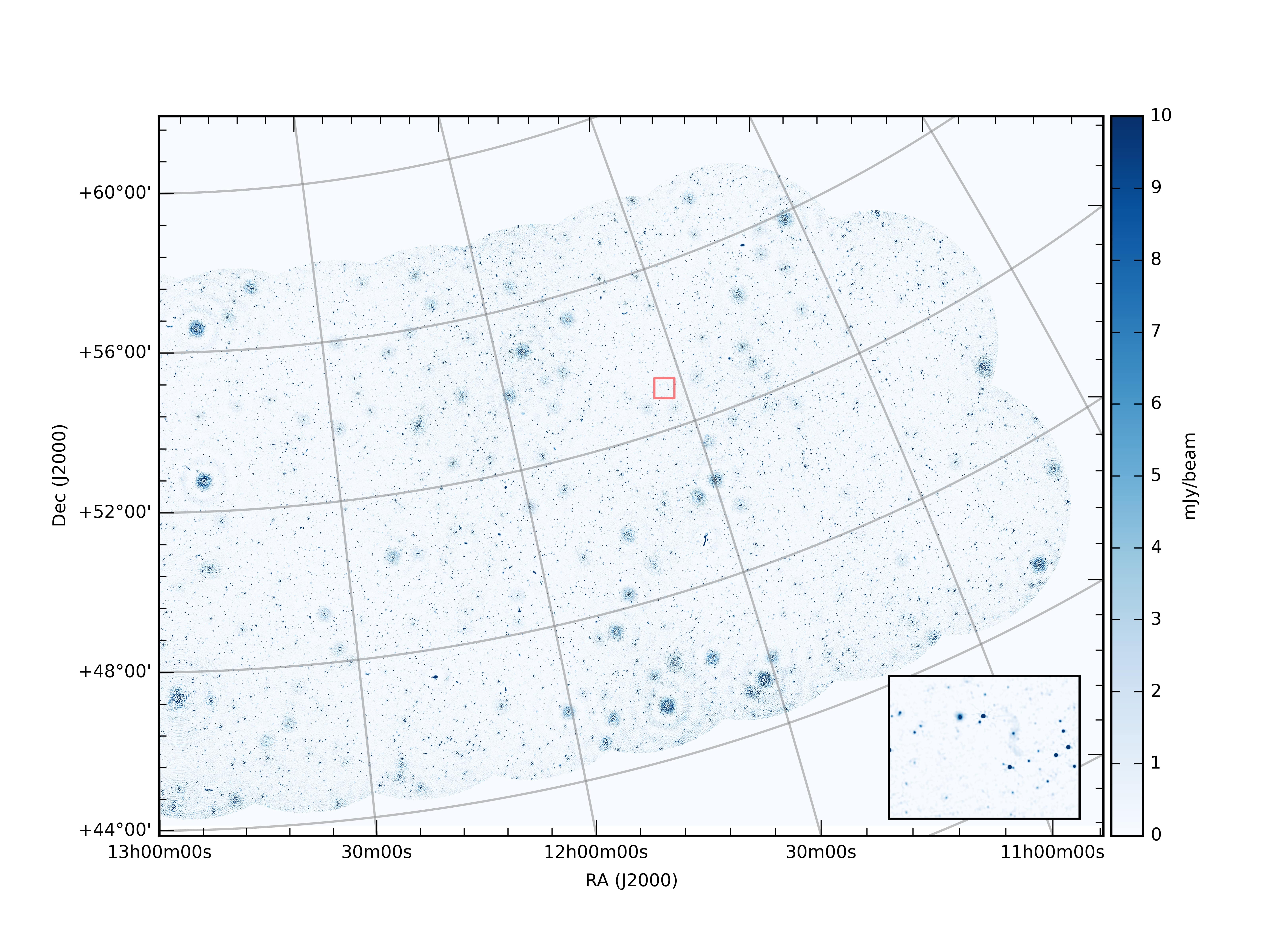}
      \label{fig:mosaic-halfA}
      \vspace{-0.8cm}
    {{Figure 16:} The western half of the HETDEX Spring Field. {A $0.5^\circ \times 0.5^\circ$ image of the region outlined in red is shown in the bottom right corner.}}
\end{sidewaysfigure*}

\begin{sidewaysfigure*}   
\centering
   \includegraphics[width=\linewidth]{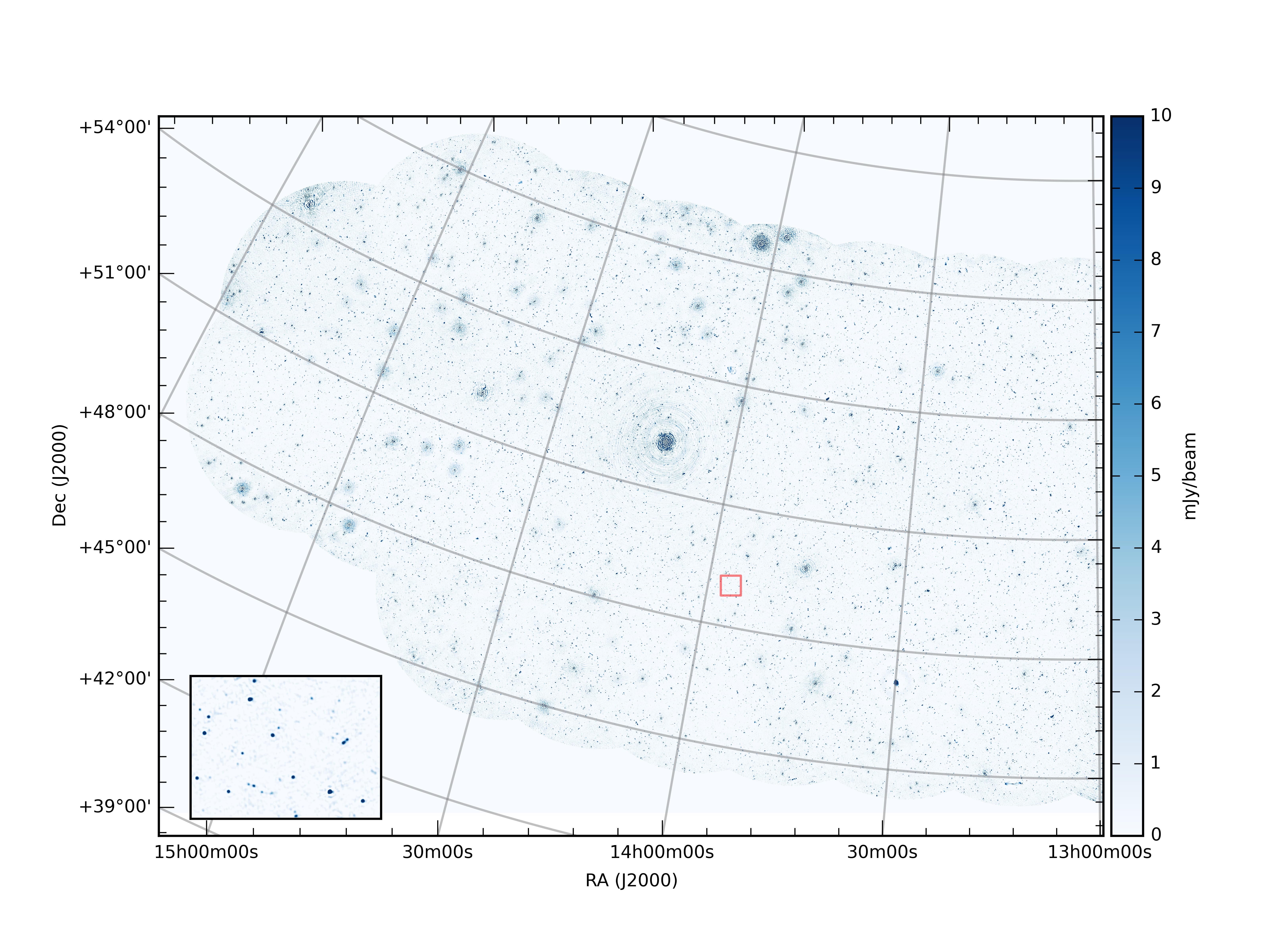} \\ 
   \label{fig:mosaic-halfB}
   \vspace{-0.8cm}
   {{Figure 17:} The eastern half of the HETDEX Spring Field. {A $0.5^\circ \times 0.5^\circ$ image of the region outlined in red is shown in the bottom left corner.}}
\end{sidewaysfigure*}

\section{Source catalogues}
\label{sec:sourcecatalogues}

Source detection on the mosaics that are centred on each pointing was performed with PyBDSM. In an effort to minimise contamination from artefacts, the catalogue was made using a conservative 7$\sigma$ detection threshold. Furthermore, as our artefacts are predominantly in regions surrounding bright sources, we utilised the PyBDSM functionality to decrease the size of the box used to calculate the local noise when close to bright sources, which has the effect of increasing the estimated noise level in these regions. Our catalogues from each mosaic are merged to create a final catalogue of the entire HETDEX Spring Field region. During this process we remove multiple entries for sources by only keeping sources that are detected in the mosaic centred on the pointing that the source is closest to the centre of. 

In the catalogue we provide the type of source, where we have used PyBDSM to distinguish isolated compact sources, large complex sources, and sources that are within an island of emission that contains multiple sources. In addition, we have attempted to distinguish between sources that are resolved and unresolved in our images. An approach that is often used to assess whether sources are resolved is to inject a distribution of point-like sources and measure the integrated flux density to peak brightness ratio as a function of signal-to-noise. An envelope can then be fitted to this distribution and real sources that are detected within the envelope can be classified as unresolved, whereas real sources outside the envelope can be classified as resolved. However, using such an approach to accurately estimate whether sources are resolved in these preliminary data release images is challenging because sources are blurred due to the uncorrected direction dependent phase errors. Rather than attempting to incorporate the phase errors into a simulation we instead define an envelope using real sources that we assume are unresolved. The population of sources that we assume are unresolved in the LOFAR images are those that correspond to entries in the FIRST catalogue that have maximum extensions of less than 5$\arcsec$. In Figure \ref{fig:source-size-envelope} we show the distribution of sources in the preliminary data release catalogue which indeed indicates that compact FIRST sources are generally still compact in the low frequency preliminary data release images. The envelope that encompasses 95\% of the sources that are compact in FIRST is described by $\rm{\frac{S_{int}}{S_{peak}} = 1.50+1.78\big(\frac{S_{peak}}{RMS}\big)^{-0.78}}$ and we use this envelope to distinguish unresolved and resolved sources in this preliminary release catalogue. We note that the median ratio of integrated flux density to peak brightness was found to vary between observations due to varying ionospheric conditions and this ratio will therefore also vary throughout the mosaiced region (see Section \ref{sec:identifying-bad-ionosphere}). This variation has not been taken into account in our classification of resolved and unresolved sources but, at all signal-to-noise ratios, the envelope that we have used to identify resolved sources is at a significantly larger integrated flux density to peak brightness ratio than the median values of this ratio for any pointings (which range from 1.08 to 1.33).

The statistical errors on the RA and Dec that are calculated by PyBDSM are smaller than those we have measured by comparing our LOFAR catalogues with the FIRST catalogue (see Section \ref{sec:astrometry}). Hence, in the catalogue we have added 1.7$\arcsec$ in quadrature with the PyBDSM statistical errors to provide more accurate error estimates. Similarly, in Section \ref{sec:flux-comparison}, we estimated that our flux measurements are accurate to 20\% and we add this in quadrature with the statistical errors provided by PyBDSM for both the peak brightness and integrated flux density measurements. Furthermore, we note that there are some very extended sources within the mosaiced region (such as the nearby galaxies M\,106 and M\,51), and for such sources our automated source finding pipeline may not accurately recover the full extent or  integrated flux density of the complex emission.  In Table 3 we show example sources from the catalogue.

In Figure \ref{fig:mosaic-noise} we show the estimated rms noise levels of the mosaiced images that were used to create the source catalogs. Within the 381 square degree region encompassed by the FWHM of the mosaiced pointings we find that the noise level is below 0.5\,mJy/beam and 1.0\,mJy/beam in 54\% and 91\% of the mosaiced region respectively. We also estimate the completeness of the catalogues following the procedure outlined in \cite{Heald_2015}. In this procedure, we first create residual mosaic images by using PyBDSM to remove the detected sources from the mosaic images that were used during the creation of the final catalog. These residual images accurately describe the properties of the mosaic images and the variation in noise across them.  A population of simulated point sources drawn from a power-law flux density distribution ($\frac{dN}{dS} \propto S^{-1.6}$) with a flux density range between 0.5\,mJy and 10\,Jy was then injected into residual mosaic images at random positions. We then attempted to detect the simulated sources on each mosaic image using the same PyBDSM settings as were used to create the final catalogue. A source was classified is detected if it is found to be within 15$\arcsec$ of its input position and with a recovered flux density that is within 10 times the error on the recovered flux density from the simulated value.  We found that sources with flux densities below 2.5\,mJy/beam were rarely detected but sources brighter than 8\,mJy/beam were detected over 90\% of the time. For a statistically robust measurement of the completeness this procedure of injecting and searching for simulated sources was repeated 50 times for each mosaic where each time 1,000 sources were injected into the image. The final completeness over the entire mosaiced region, which is the fraction of recovered sources as a function of flux density, is shown in Figure \ref{fig:mosaic-noise}. We find that the catalogue is 50\% complete over 1.1\,mJy and 90\% complete over 3.9\,mJy.

\begin{figure}   
\centering
   \includegraphics[width=\linewidth]{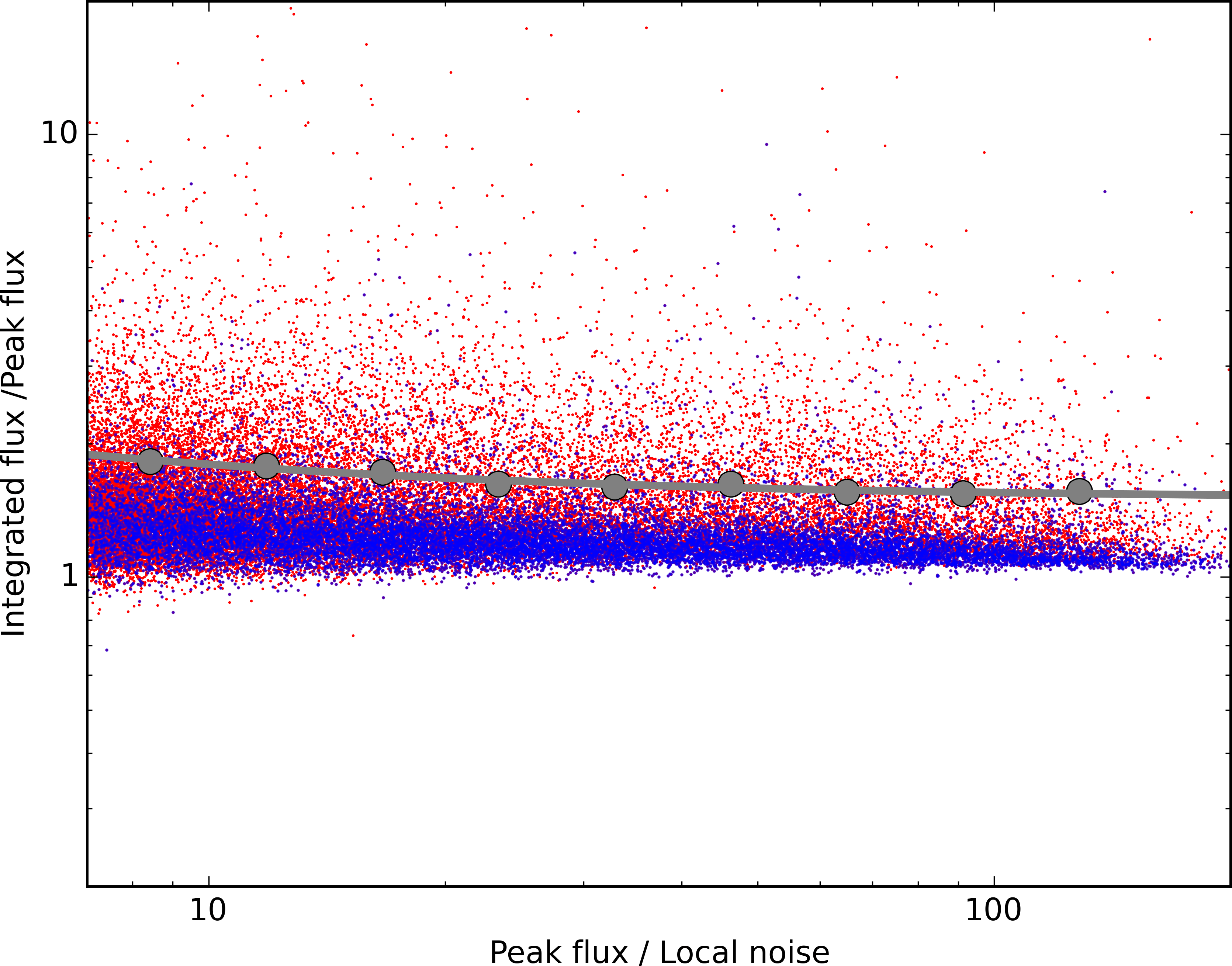}
    \caption{The ratio of the integrated flux density to peak brightness for sources in the preliminary data release catalogue as a function of the signal-to-noise ratio. Sources with a size of less than $5\arcsec$ in the FIRST catalogue are shown in blue and all other sources are shown in red. The large dots indicate the boundary that contains 95\% of the compact FIRST detected sources and the line shows the best fit to this boundary which is given by $\rm{\frac{S_{int}}{S_{peak}} =1.50+1.78\big(\frac{S_{peak}}{RMS}\big)^{-0.78}}$.}
   \label{fig:source-size-envelope}
\end{figure}

\begin{figure*}   
\centering
   \includegraphics[height=6cm]{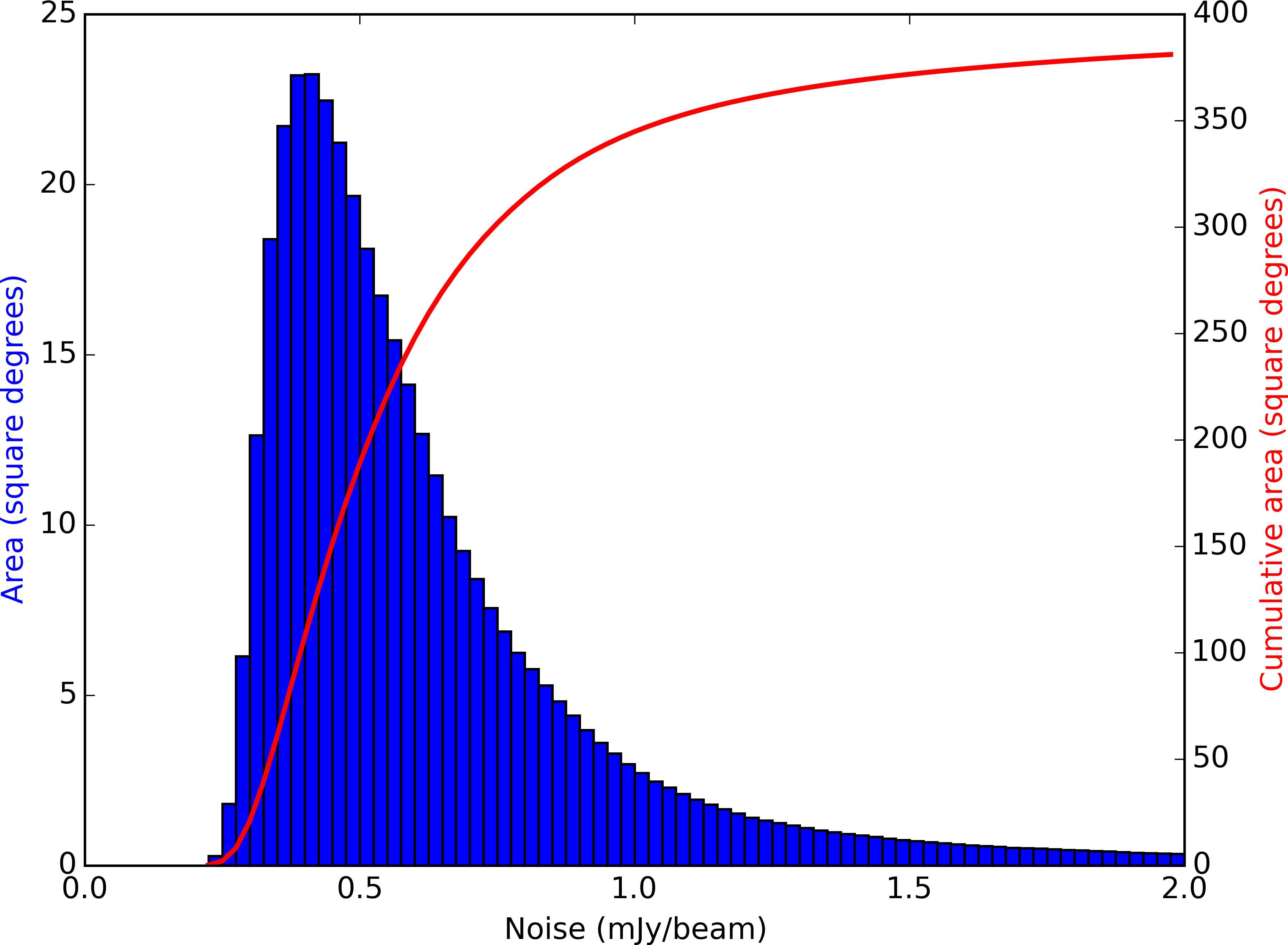}
   \hspace{0.5cm}
   \includegraphics[height=6cm]{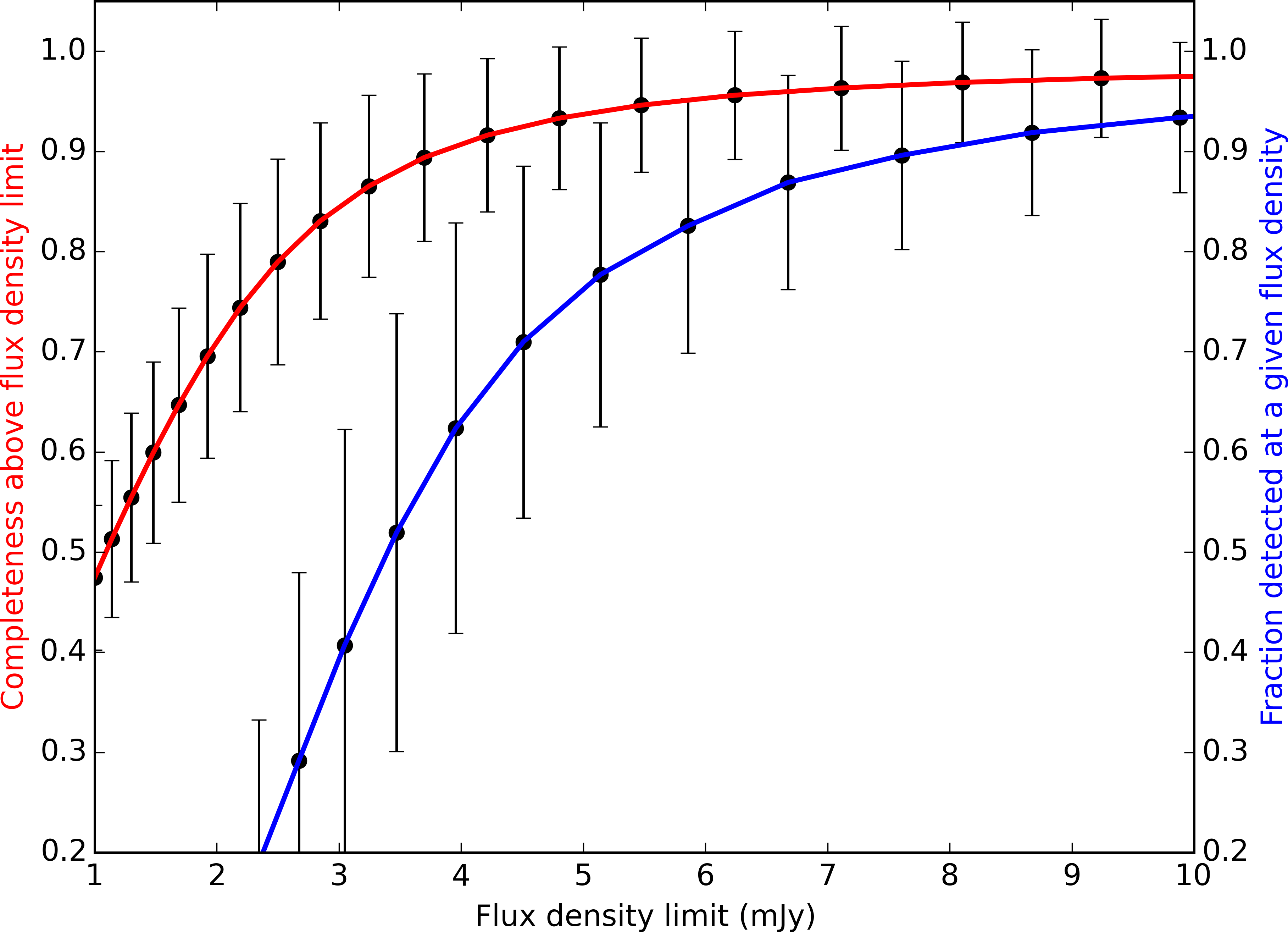}
    \caption{Left: The estimated noise variations on the direction-independent calibrated mosaiced images within the 381 square degree region encompassed by the FWHM of the mosaiced pointings. The red line shows the cumulative area of the mosaiced region that has a estimated noise less than a given value. The histogram shows the distribution of noise estimates within the  mosaiced region. Right: The estimated {cumulative completeness} of the preliminary data release catalogue {(red)} and the fraction of simulated sources that are detected as a function of flux density (blue). {A large number of sources were injected during the completeness simulations and as a consequence the poissonian errors are negligible. However, the spatial variation in noise is substantial and the error bars show the standard deviation of the measurements as a function of position.}}
   \label{fig:mosaic-noise}
\end{figure*}

\begin{sidewaystable*}
\centering
{{Table 3:} An example of entries in the source catalogue for the 54 direction-independent calibrated images. The entire catalogue contains over 44,000 sources. The entries in the catalogue are: source identifier (ID), J2000 right ascension (RA), J2000 declination (Dec), peak brightness ($\rm{S_{peak}}$), integrated flux density ($\rm{S_{int}}$),  a flag indicating whether the source is resolved (R) or unresolved (U), the local noise at the position of the source (RMS noise), the type of source (where `S' indicates an isolated source which is fit with a single gaussian; `C' represents sources that are fit by a single gaussian but are within an island of emission that also contains other sources; and `M' is used for sources which are extended and fitted with multiple gaussians), and the mosaic identifier. The right ascension, declination, peak brightness and integrated flux density are assigned both a formal error from the source fitting and a total error that accounts for the 1.7$\arcsec$ astrometric uncertainty and the 20\% flux uncertainty.}\\
\label{tab:catalogue-example}
\begin{tabular}{lccccccccccccccccc}
\hline
Source ID & RA & $\sigma_{RA}$ & DEC & $\sigma_{DEC}$ & $S_{peak}$ & $\sigma_{Speak}$  & $S_{int}$ & $\sigma_{Sint}$ & Resolved & RMS  & Type & Mosaic \\  
 &  & ($\sigma_{RA,tot}$) &  & ($\sigma_{DEC,tot}$) & & ($\sigma_{Speak,tot}$) &   &  ($\sigma_{Sint,tot}$) &   &  noise &  & ID \\ 
  & ($^\circ$)  & ($\arcsec$) & ($^\circ$)  & ($\arcsec$) & (mJy/beam) & (mJy/beam) & (mJy)  &  (mJy) &  & (mJy) &  &  &   \\  \hline
ILTJ104322.7+473446.5 & 160.8445 & 0.7 (1.8) & 47.5796 & 0.7 (1.8) & 8.0 & 0.5 (1.7) & 9.2 & 0.8 (2.0) & U & 0.5 & C & P3  \\
ILTJ105059.3+515054.5 & 162.7473 & 1.7 (2.4) & 51.8485 & 1.6 (2.3) & 3.4 & 0.4 (0.8) & 5.2 & 0.7 (1.2) & U & 0.4 & S & P4  \\
ILTJ110856.6+480024.4 & 167.2356 & 0.6 (1.8) & 48.0068 & 0.5 (1.8) & 17.0 & 0.5 (3.4) & 38.0 & 0.7 (7.6) & R & 0.5 & S & P7  \\
ILTJ123820.5+533450.5 & 189.5856 & 0.9 (1.9) & 53.5807 & 1.1 (2.0) & 16.7 & 1.5 (3.6) & 17.5 & 2.5 (4.3) & U & 1.5 & S & P29  \\
ILTJ112629.2+532151.7 & 171.6216 & 1.2 (2.1) & 53.3644 & 1.7 (2.4) & 3.3 & 0.5 (0.8) & 3.4 & 0.8 (1.0) & U & 0.5 & S & P12  \\
ILTJ115132.5+471429.2 & 177.8854 & 1.4 (2.2) & 47.2414 & 2.0 (2.6) & 9.0 & 1.2 (2.2) & 11.5 & 2.0 (3.1) & U & 1.2 & S & P15  \\
ILTJ124209.0+555428.1 & 190.5377 & 4.9 (5.1) & 55.9078 & 2.0 (2.6) & 2.8 & 0.4 (0.7) & 13.0 & 0.5 (2.6) & R & 0.4 & S & P191+55  \\
ILTJ124003.6+574429.8 & 190.0152 & 1.4 (2.2) & 57.7416 & 1.2 (2.1) & 11.3 & 1.3 (2.6) & 14.1 & 2.0 (3.5) & U & 1.2 & S & P191+55  \\
ILTJ132112.1+574121.3 & 200.3005 & 2.1 (2.7) & 57.6892 & 2.5 (3.0) & 5.3 & 0.9 (1.4) & 8.2 & 1.4 (2.1) & U & 0.8 & C & P200+55  \\
ILTJ151613.0+505206.5 & 229.0541 & 0.7 (1.9) & 50.8685 & 0.8 (1.9) & 6.5 & 0.5 (1.4) & 6.8 & 0.8 (1.6) & U & 0.5 & S & P227+50  \\
ILTJ150730.3+501951.5 & 226.8761 & 0.8 (1.9) & 50.3310 & 1.0 (2.0) & 4.4 & 0.4 (1.0) & 4.1 & 0.7 (1.1) & U & 0.4 & S & P227+50  \\
ILTJ150506.2+505458.1 & 226.2760 & 0.7 (1.8) & 50.9161 & 0.8 (1.9) & 8.9 & 0.6 (1.9) & 10.1 & 0.9 (2.2) & U & 0.6 & S & P227+50  \\
ILTJ142842.4+481322.7 & 217.1767 & 1.1 (2.0) & 48.2230 & 1.3 (2.2) & 5.0 & 0.5 (1.1) & 6.8 & 0.8 (1.6) & U & 0.5 & S & P217+47  \\
ILTJ140150.1+460845.3 & 210.4587 & 1.6 (2.3) & 46.1459 & 1.6 (2.3) & 3.3 & 0.4 (0.8) & 4.3 & 0.7 (1.1) & U & 0.4 & S & P210+47  \\
ILTJ135725.8+484433.4 & 209.3577 & 1.3 (2.1) & 48.7426 & 1.3 (2.1) & 3.0 & 0.3 (0.7) & 4.0 & 0.5 (0.9) & U & 0.3 & S & P210+47  \\
ILTJ143733.8+563232.0 & 219.3909 & 0.8 (1.9) & 56.5422 & 0.4 (1.8) & 16.8 & 0.6 (3.4) & 37.0 & 0.8 (7.5) & R & 0.6 & S & P218+55  \\
ILTJ151911.5+512419.0 & 229.7978 & 1.9 (2.6) & 51.4053 & 2.5 (3.0) & 4.3 & 0.6 (1.1) & 7.7 & 0.9 (1.8) & U & 0.6 & S & P227+53  \\
ILTJ150703.0+540008.3 & 226.7625 & 0.9 (1.9) & 54.0023 & 0.9 (1.9) & 7.6 & 0.6 (1.6) & 10.6 & 0.9 (2.3) & U & 0.5 & S & P227+53  \\
ILTJ132723.5+464943.9 & 201.8478 & 0.8 (1.9) & 46.8289 & 0.8 (1.9) & 6.2 & 0.4 (1.3) & 7.8 & 0.7 (1.7) & U & 0.4 & S & P39  \\
ILTJ111612.8+555350.7 & 169.0535 & 0.5 (1.8) & 55.8974 & 0.6 (1.8) & 8.9 & 0.4 (1.8) & 10.5 & 0.7 (2.2) & U & 0.4 & S & P169+55  \\
ILTJ122630.6+531346.8 & 186.6274 & 0.5 (1.8) & 53.2297 & 0.5 (1.8) & 8.8 & 0.4 (1.8) & 10.1 & 0.6 (2.1) & U & 0.3 & S & P25  \\ \hline  
 \end{tabular}
\end{sidewaystable*}

\section{Public data release}
\label{sec:publicddatarelease}

The images and catalogues that were presented in this paper have now been made publicly available\footnote{http://lofar.strw.leidenuniv.nl}. This released dataset consists of direction-independent calibrated images of 54 pointings in the region from right ascension 10h45m00s to 15h30m00s and declination 45$^\circ$00$\arcmin$00$\arcsec$ to 57$^\circ$00$\arcmin$00$\arcsec$ (63 pointings were calibrated but two had a very high fraction of flagged data and seven were found to have poor ionospheric conditions during the observation). We have also released a catalogue of the region which contains over 44,000 sources which were detected with a signal in excess of seven times the local noise on the mosaiced images. The sensitivity within the 381 square degrees that was mosaiced varies significantly (see Figure \ref{fig:mosaic-noise}) but we have estimated that the catalogue is 90\% complete for sources with flux densities in excess of 3.9\,mJy/beam.
 
\section{Direction-dependent calibration}
\label{sec:facet-calibration}

Whilst the images presented in this publication are sensitive low-frequency images, these LOFAR datasets, if accurately calibrated, will produce high fidelity images with $\approx5\arcsec$ resolution and $\approx100\,\mu$Jy/beam sensitivity as was demonstrated by e.g. \cite{vanWeeren_2015a}, \cite{vanWeeren_2015b}, \cite{Shimwell_2016}, \cite{Williams_2015} and \cite{Hardcastle_2016}. Routinely producing such images is the challenge that the LOFAR surveys team is presently tackling. We are putting in place strategies to deal with the large data rate, the computational expense of the calibration, the manual interaction of the calibration, and how to effectively share the data for maximum scientific exploitation. In a future data release, we intend to make direction-dependent calibrated images and catalogues available to the wider scientific community. 

As a qualitative demonstration of the improvement that direction-dependent calibration will offer, in Figure \ref{fig:facet-comparison} we show 120-168\,MHz images of one of our datasets before and after facet calibration\footnote{The facet calibration code can be found at https://github.com/lofar-astron/factor}. The difference in noise level, resolution, dynamic range and image fidelity is clear. The noise measured in the same region of both images is 360\,$\mu$Jy/beam and 100\,$\mu$Jy/beam for the direction-independent and dependent calibrated images respectively. The resolution of the direction-independent calibrated image is $25\times25\arcsec$ whereas the direction-dependent calibrated image has a resolution of $4.8\times7.9\arcsec$. For a detailed evaluation of the quality of typical facet calibrated images we refer the reader to \cite{Williams_2015} and \cite{Hardcastle_2016}. 

\begin{figure*}   \centering
   \includegraphics[width=0.49\linewidth]{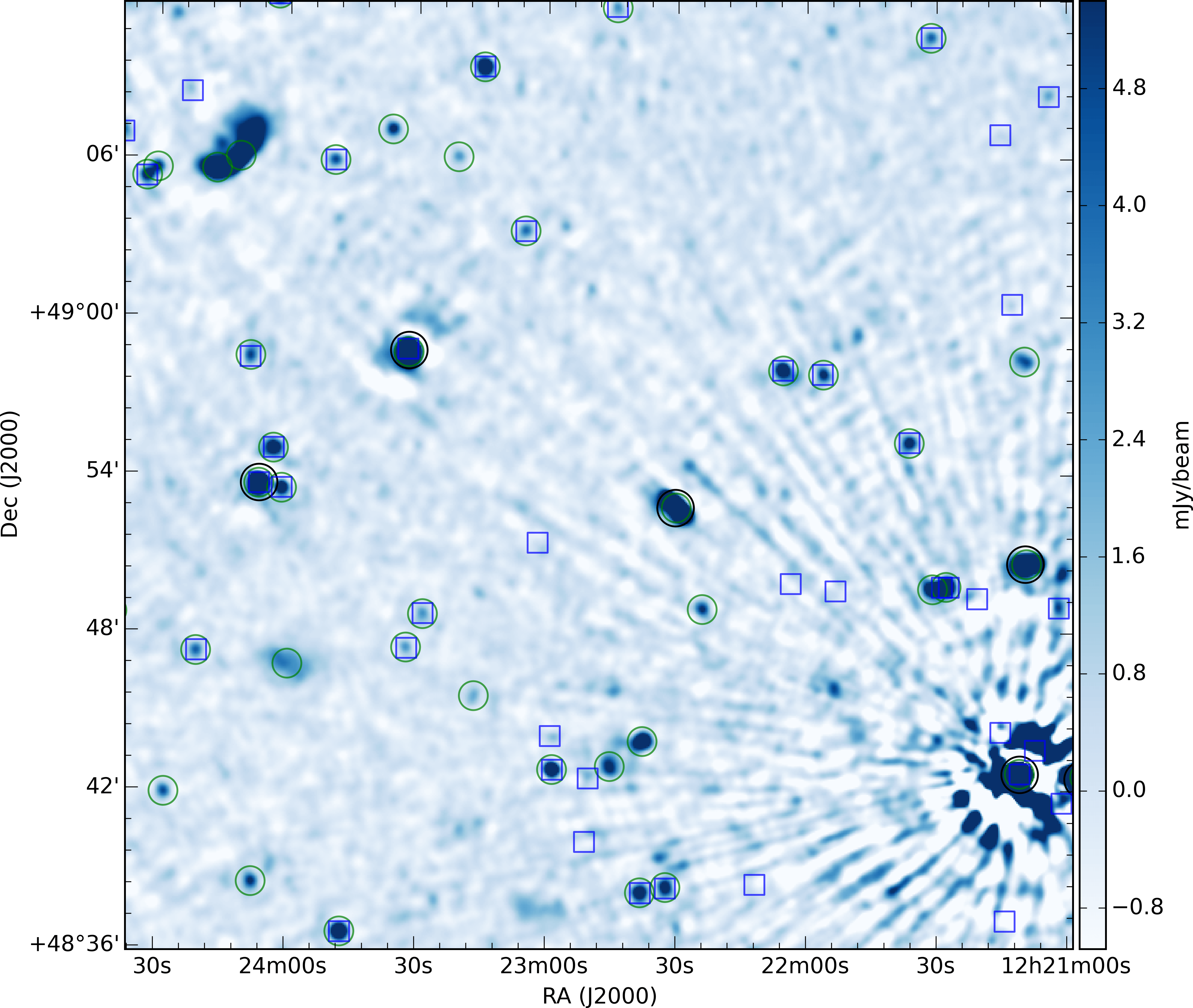} 
   \includegraphics[width=0.49\linewidth]{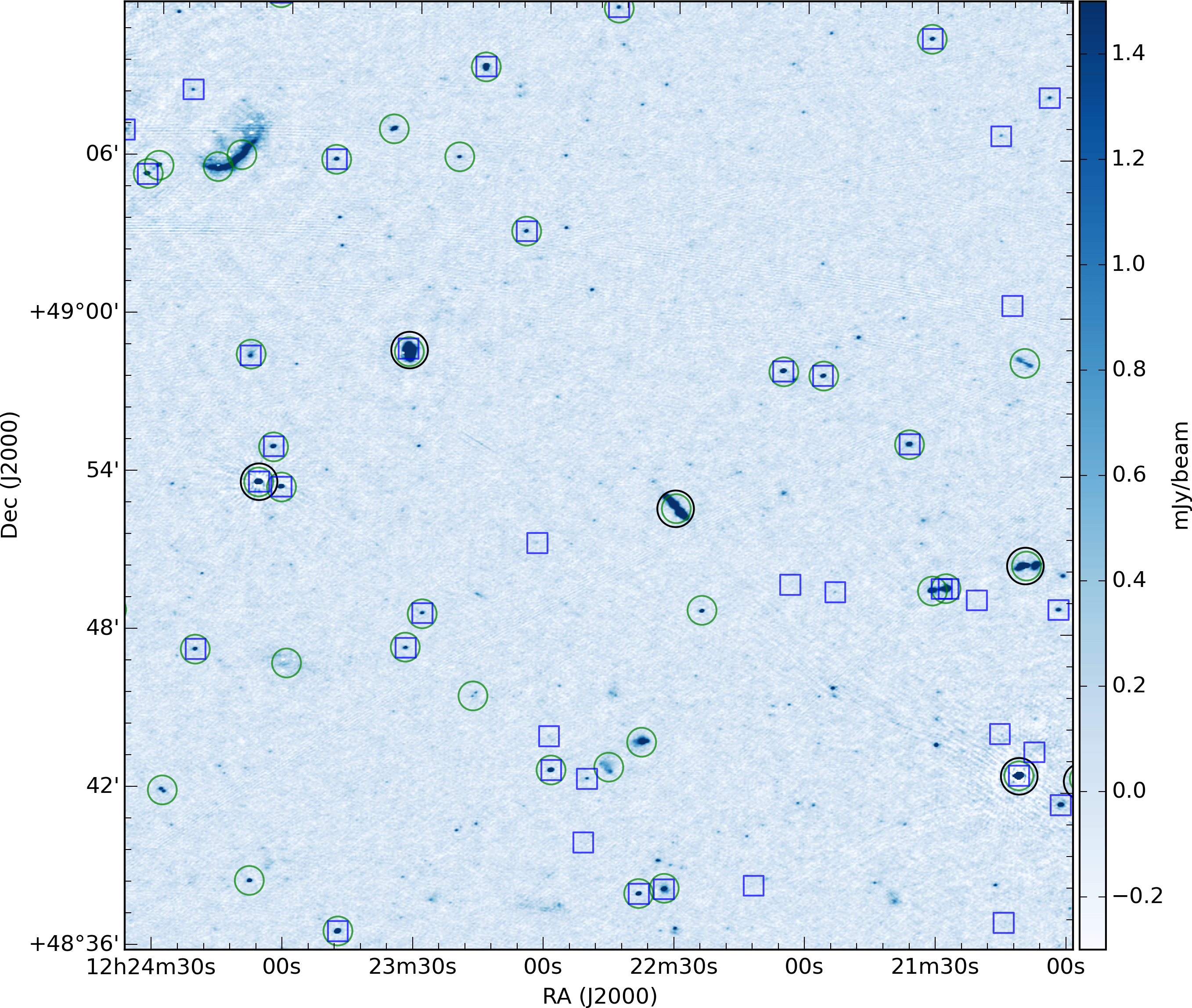} 
   \caption{A comparison between the direction-independent calibrated images and those we expect from direction-dependent calibration. On the right the facet calibration method has been used to create a high resolution, high fidelity, approximately thermal noise limited image and on the left is the same region in our direction-independent calibrated images. The colour scales on both images is between $-3$ and 10 times the noise where the direction-independent calibrated image noise is 360\,$\mu$Jy/beam and the facet calibrated image noise is 100\,$\mu$Jy/beam. The green circles show the positions of sources detected at seven times the noise in the direction-independent LOFAR image. The larger black circles and the blue squares indicate entries in the TGSS and FIRST catalogues respectively.}
   \label{fig:facet-comparison}
\end{figure*}

\section{Scientific potential}
\label{sec:scientificpotential}
A detailed scientific exploitation of the LoTSS data is beyond the scope of this paper. However, below we do offer an insight into a few areas of potential scientific value.

Sensitive images have the potential to create large samples of radio sources located at high-redshift such as J1429+544, which is a $z=6.21$ quasar (\citealt{Willott_2010}) that is well detected in our LOFAR images with a peak brightness of 9\,mJy/beam (see Figure \ref{fig:science-examples}). For example, in Figure \ref{fig:specz-examples} we show the magnitude-redshift plane for BOSS radio-detected quasars (\citealt{Paris_2014}) inside the footprint of the LOFAR images we have released. This figure demonstrates that in our preliminary LOFAR images we detect 35\% more radio quasars than FIRST. Moreover, combining these radio observations with dropout searching techniques that are based on the identification of sources with very red optical/near-infrared is an effective way to eliminate stellar contaminants from photometric samples and increase the success rate for spectroscopic follow-ups (e.g. \citealt{McGreer_2009} and \citealt{Banados_2015}). These searches will eventually identify many powerful radio-sources at $z>6$, which are ideal targets to carry out \ion{H}{I} 21cm absorption line studies in the Epoch of Reoinization. {The detection of the \ion{H}{I} 21cm line will allow us to study the immediate AGN surroundings and interstellar gas in the host galaxy, constrain the cosmic evolution of gas excitation, and detect possible homogeneity of the last neutral regions from cosmic reionization (e.g. \citealt{Carilli_2002}, \citealt{Carilli_2007} and \citealt{Furlanetto_2002}). }

The LoTSS images can be used to examine the propagation of cosmic ray electrons in nearby galaxies such as M\,106 and M\,51 (Figure \ref{fig:science-examples}). For example, \cite{Mulcahy_2014} and \cite{Mulcahy_2016} used similar observations to reveal synchrotron emission from a highly extended disk of old low-energy electrons that have propagated out to a radius of 16\,kpc from the centre of the grand-design spiral galaxy M\,51. A comparison of scale lengths at low and high frequencies as well as the scale-dependent radio to far infrared correlations at low and high frequencies gave clear evidence for the propagation of cosmic rays by {diffusion. A} similar study is being performed to characterise the low-frequency emission in the nearby spiral galaxy M\,106 (\citealt{Sridhar_2016}). {This galaxy hosts distinctive anomalous radio arms (e.g \citealt{Courtes_1961} and \citealt{vanderkruit_1972}) but their precise location with respect to the star forming disk has remained a matter of debate (see e.g. \citealt{Wilson_2001}). A reprocessing of the LoTSS data to correct it for the ionospheric Faraday Rotation (see \citealt{Cameron_2016} for details) can provide polarisation measurements that will help pinpoint the location of the anomalous arms.} In addition, the continuum images that have excellent surface brightness sensitivity will be used to examine the old cosmic ray population and constrain the magnetic field strength of the anomalous arms and the entire star forming disk. 

In approximately 100 galaxy clusters diffuse steep spectrum synchrotron emission that is associated with the intra-cluster medium (ICM) has been observed (see \citealt{Ferrari_2008}, \citealt{Bruggen_2012}, \citealt{Feretti_2012} and \citealt{Brunetti_2014} for recent reviews). The exact cause of the emission is still debated, but it is primarily classified as radio halos and radio relics and the favoured formation scenarios for these include post-merger turbulence and shock fronts respectively. LoTSS is expected to reveal many new examples of such emission and to characterise known examples in great detail (e.g. \citealt{Cassano_2010}). In the preliminary data release more than 30 massive, Sunyaev Zel'dovich detected galaxy clusters (\citealt{Planck_2015}) lie within the mapped region. This offers the opportunity for detailed studies of interesting objects and a large unbiased study to further understand the prevalence of radio halos and radio relics and which clusters they occur in. 

Two interesting examples for detailed {cluster} studies are Abell 1550 ($z=0.254$) and Abell 1682 ($z=0.226$) whose low-frequency emission we show in Figure  \ref{fig:science-examples}. Abell 1550 was previously studied by \cite{Govoni_2012} who concluded it was in a merging state after identifying an extension in the ROSAT X-ray emission as well as a displacement between the centroids of the X-ray emission and the optical galaxy {distribution; using VLA observations they also} detected diffuse radio emission from the ICM with a total 1.4\,GHz flux density of 7.7$\pm$1.6\,mJy and classified this as a radio halo. The steep spectrum ($\alpha$ is typically less than $-1$ for radio halos) implies that the total emission from the ICM should exceed $\sim$70\,mJy at {150\,MHz. In} the LoTSS preliminary data release images at 12h29m05s +47$^\circ$37$\arcmin$00$\arcsec$ there is a tentative detection of faint diffuse emission that appears to be associated with the ICM of Abell 1550, but there are several complex sources within the region of emission. {Re-processing of the LOFAR data at higher angular resolution and sensitivity would allow the diffuse emission associated with the ICM to be precisely distinguished from contaminating radio sources to confirm this} radio halo and further characterise it to provide additional insights into the dynamical state of this cluster. The galaxy cluster Abell 1682 has been well studied at radio frequencies from 150\,MHz to 1.4\,GHz by \cite{Venturi_2008}, \cite{Venturi_2011}, \cite{Venturi_2013} and \cite{Macario_2013}. {The cluster contains various regions of diffuse emission, arguably the most interesting of which is the faint emission around 13h06m56s +46$^\circ$32$\arcmin$32$\arcsec$.} This very steep spectrum ($\alpha_{240}^{610}=-2.09\pm0.15$) diffuse emission lies in a trough between two main regions of X-ray emission from the intra-cluster medium, and is thought to be either the brightest region of an underlying radio halo, a dying radio galaxy, or possibly even a radio relic (e.g. \citealt{Macario_2013}). If it is a radio halo then it falls into the category of ultra-steep radio halos and these objects, of which only a few are known, are predicted by some models describing the origin of radio halos (see \citealt{Cassano_2006} and \citealt{Brunetti_2008}). A detailed of analysis of LOFAR HBA and LBA data of Abell 1682 is being conducted to produce sensitivity high- and low-resolution images to further constrain the spectral properties of the radio emission and thoroughly assess its origin (\citealt{Clarke_2016}).

It is thought that the Square Kilometre Array (SKA) will detect approximately a million tailed radio galaxies (\citealt{Hollitt_2015}) and, similarly, LoTSS will detect a substantial number to facilitate interesting statistical studies.  However, few tailed radio galaxies are more spectacular than IC 711 ($z$=0.034) which, at a length of $\sim$1\,Mpc, is one of the longest known tailed radio galaxies (see e.g. \citealt{Vallee_1976}). Detailed studies of tailed radio galaxies like this provide a history of their motion and of the interaction between the tails and the ICM.  For example, the oldest region of the tails of IC 711 is thought to be $\sim$2\,Gyrs old, and the multiple intensity variations along the structure are thought to be caused by in situ reaccelaration, whereas the abrupt increase in the width of the tail close to its northern edge may reflect a sudden change in the jets physical conditions within the optical nucleus $\sim$1.6\,Gyrs ago or the properties of the surrounding ICM (\citealt{Vallee_1988}). The LoTSS observations of IC 711 that are presented in  Figure  \ref{fig:science-examples} are by far the most sensitive low-frequency observations of this object. A careful analysis may reveal that the jet is even longer than previously known, and by combining with higher frequency measurements the spectral index variations within the tails can be accurately mapped to help further understand the particle acceleration mechanisms within the {tails.}

Along with such examples of spectacular and peculiar sources, the survey can also be used to provide crucial insights in to the dynamics, energetics and duty-cycle of the radio galaxy population as a whole. LOFAR studies of individual active (e.g. \citealt{Orru_2015}, \citealt{Harwood_2016}, and \citealt{Heesen_2016} ), and remnant (\citealt{Shulevski_2015b} and \citealt{Brienza_2016}) radio galaxies have already expanded our understanding of radio galaxies at low frequencies, but studies of the larger sample of sources that we have imaged will provide the opportunity to determine the applicability of these findings to the population as a whole. 

\begin{figure}   
\centering
   \includegraphics[width=\linewidth]{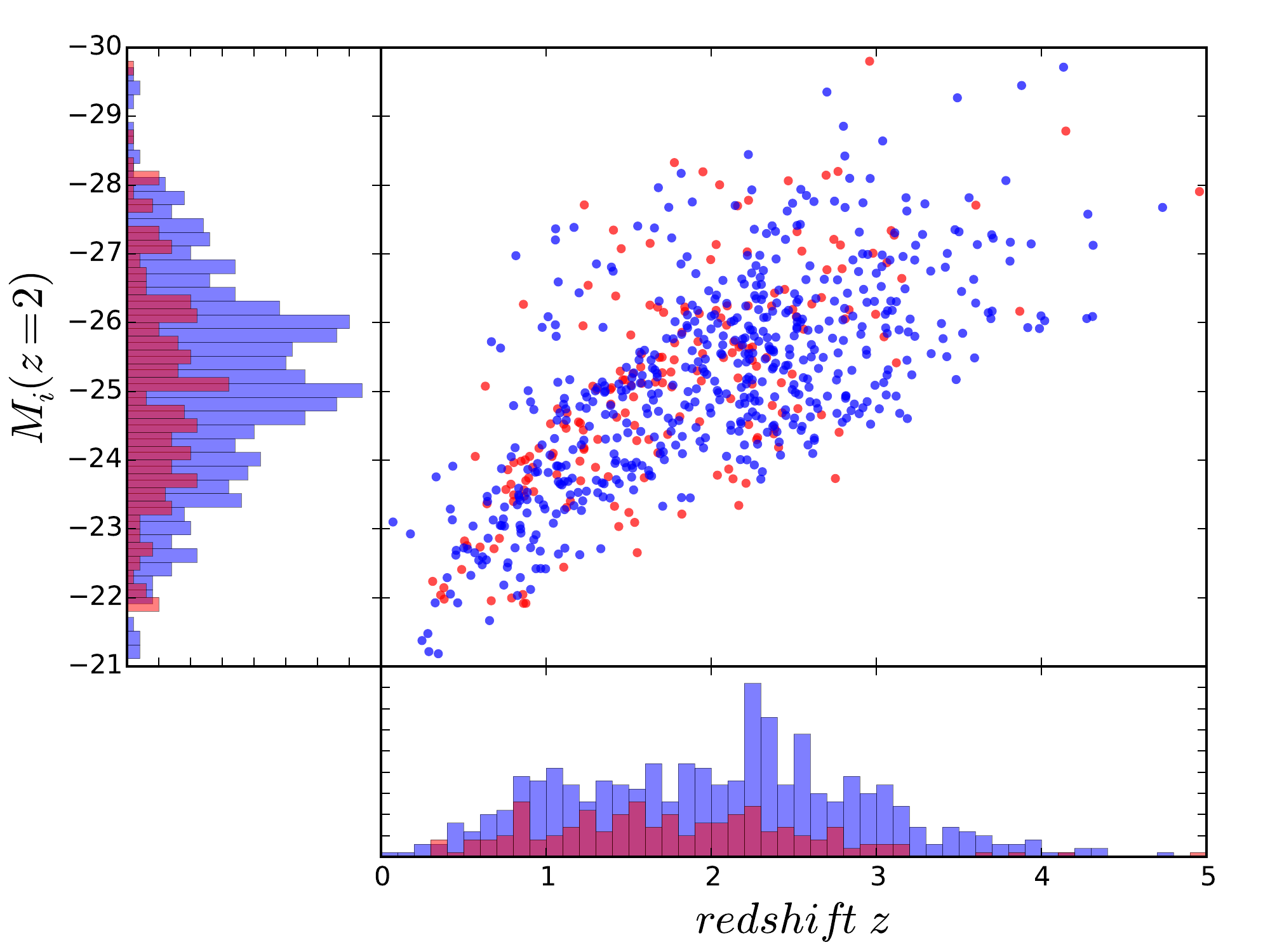}
   \caption{The distribution of quasars in the magnitude-redshift space for the BOSS radio-loud quasars (\citealt{Paris_2014}) inside the survey footprint. Whilst 551 radio-quasars are detected in both the FIRST and LOFAR images (blue circles), LOFAR is able to detect a further 191 that are not detected in FIRST (red circles). The absolute magnitude in the i band at $z = 2$  $\rm{M_i}(z=2)$ is calculated using the K-correction from \protect\cite{Richards_2006}. The left and bottom panels show the $\rm{M_i}(z=2)$ and redshift histograms.}
   \label{fig:specz-examples}
\end{figure}

\begin{figure*}   
\centering
   \includegraphics[width=0.45\linewidth]{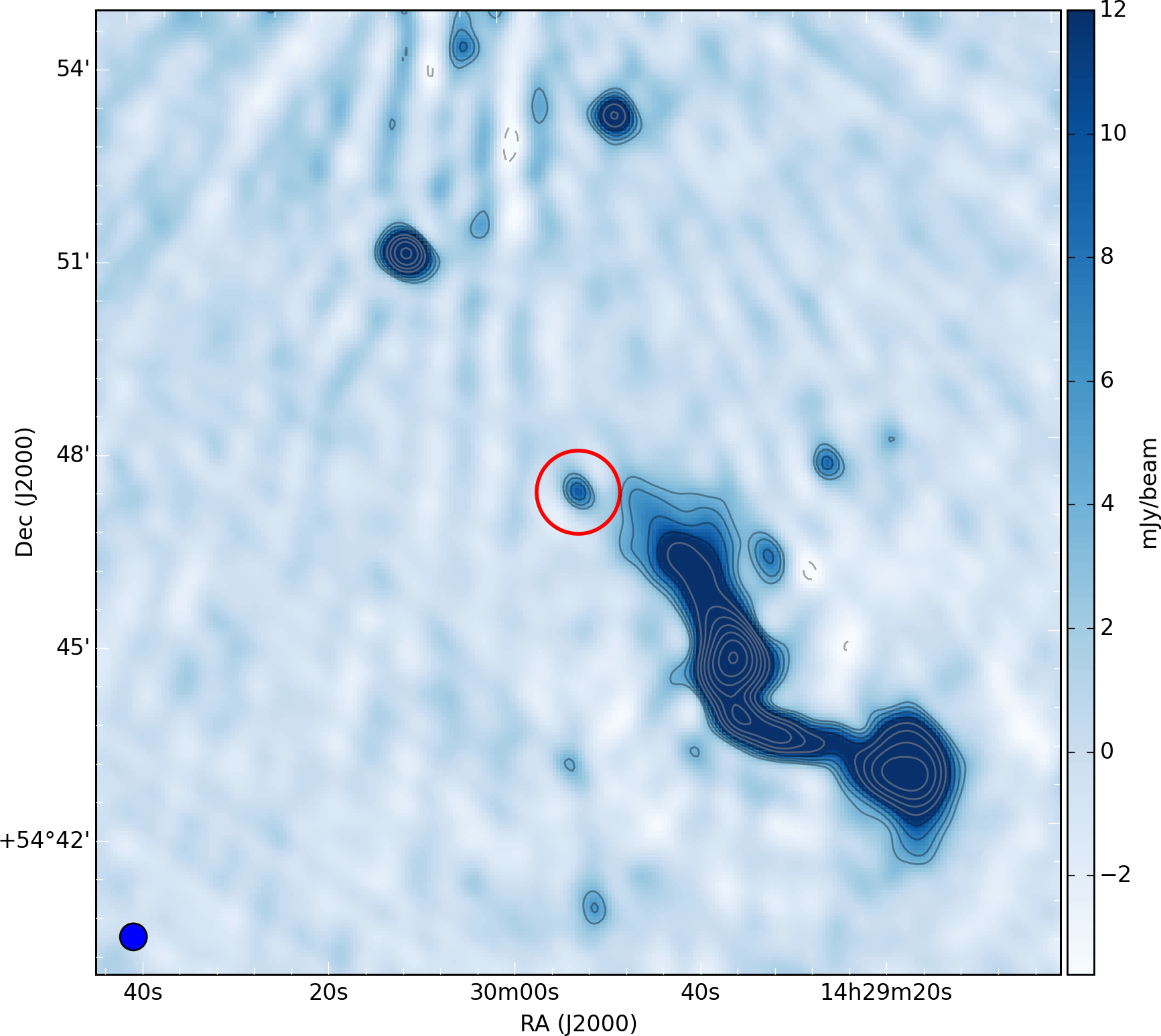} 
   \includegraphics[width=0.45\linewidth]{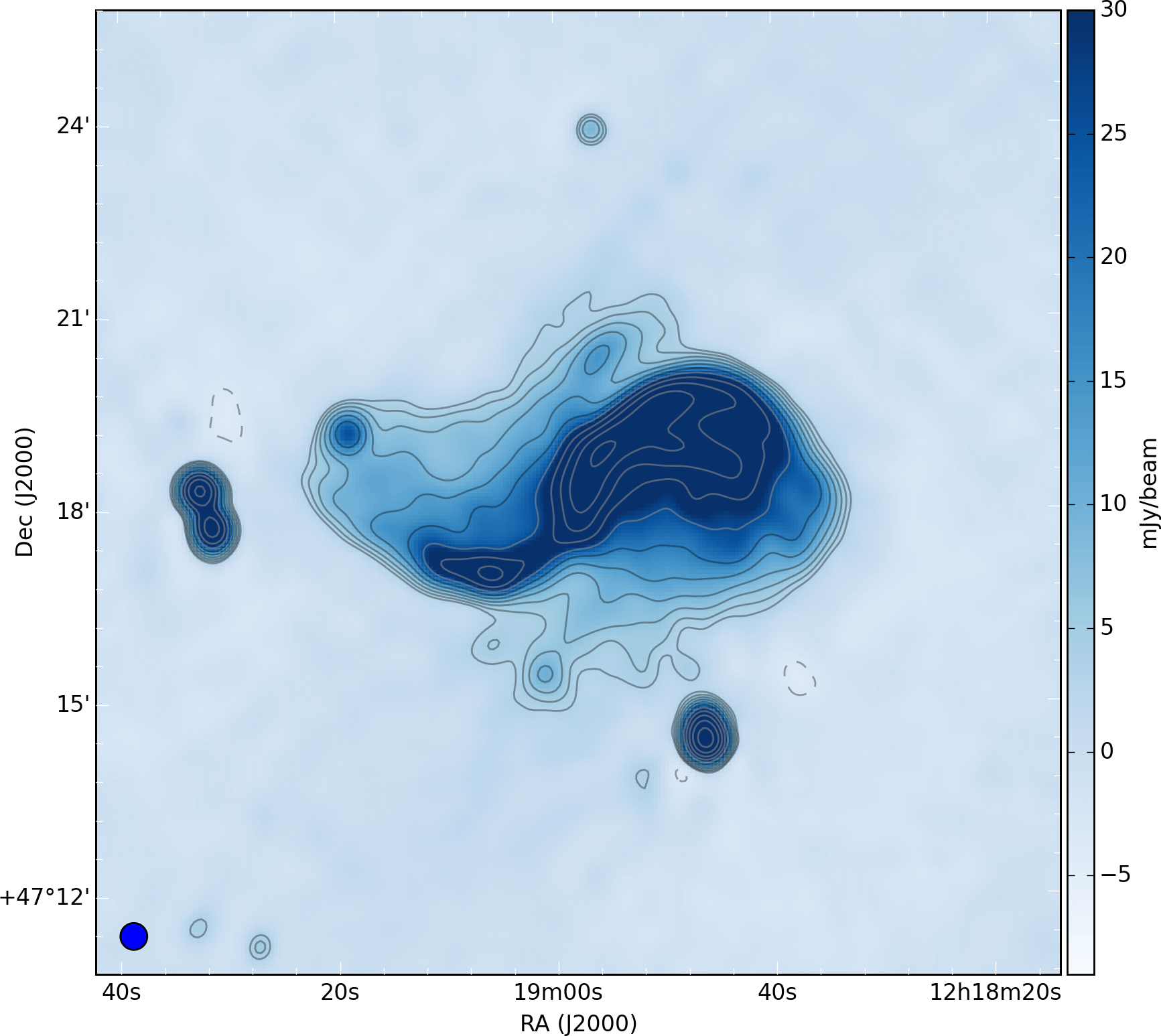}
   \includegraphics[width=0.45\linewidth]{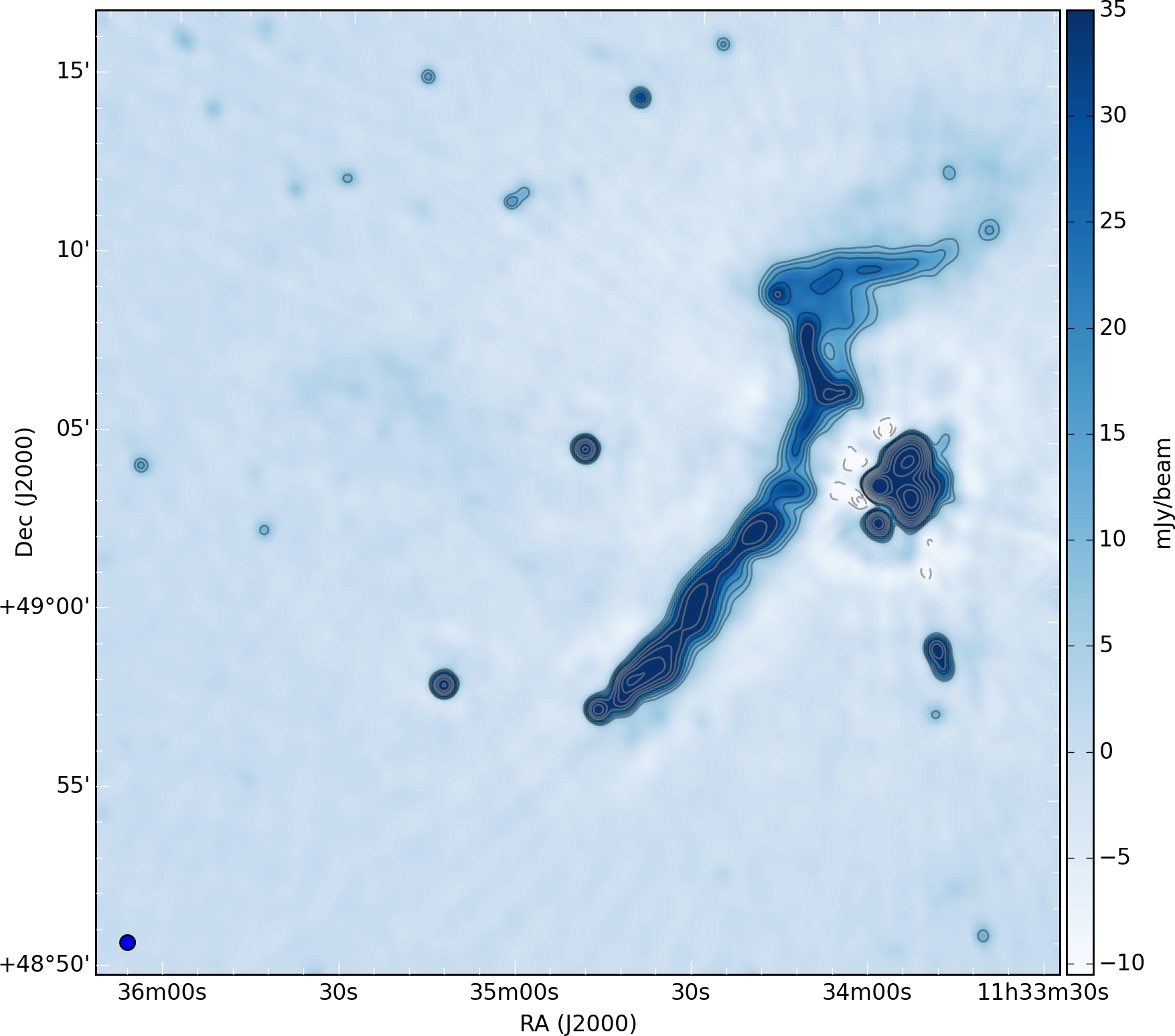}
    \includegraphics[width=0.45\linewidth]{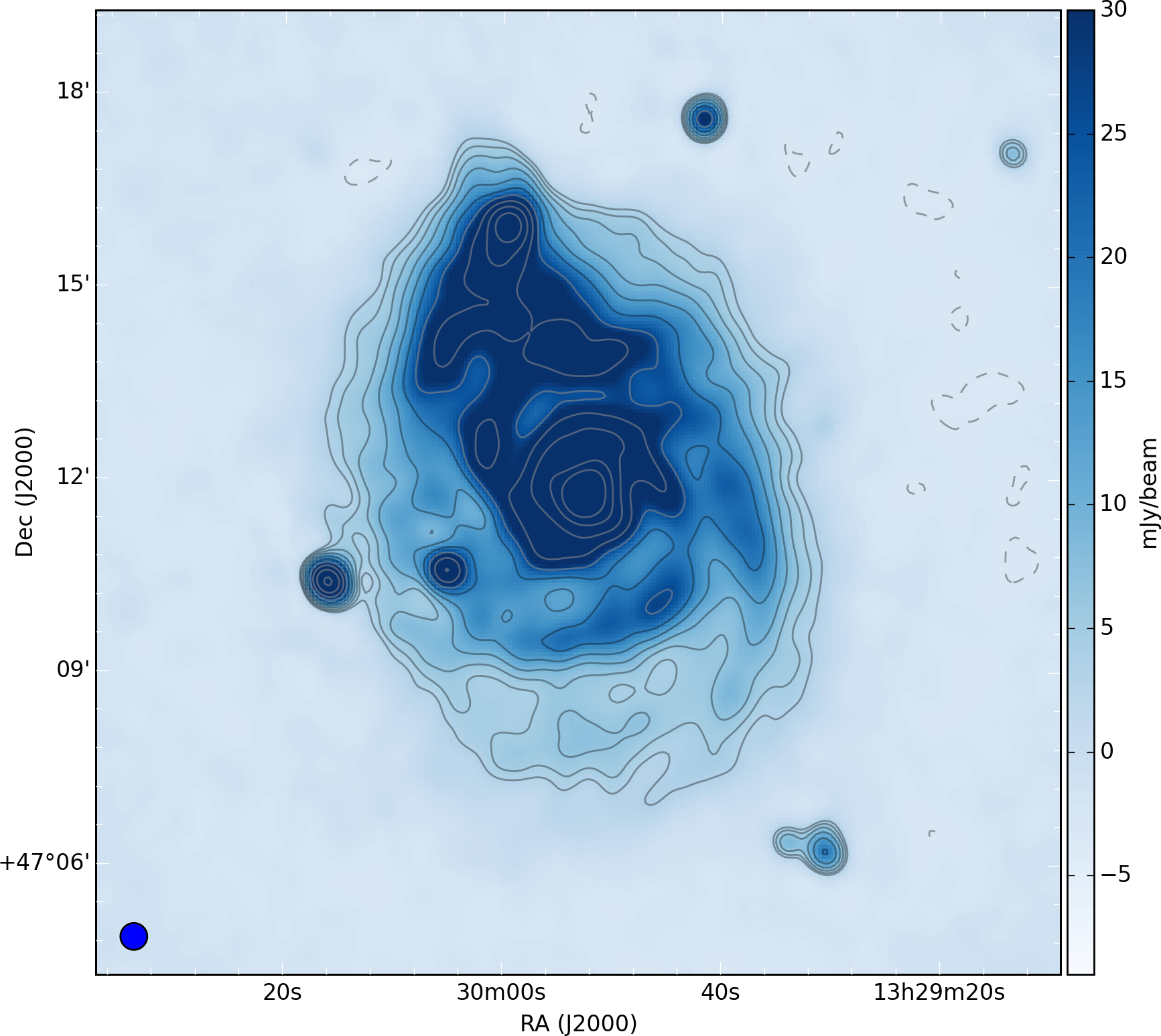} 
    \includegraphics[width=0.45\linewidth]{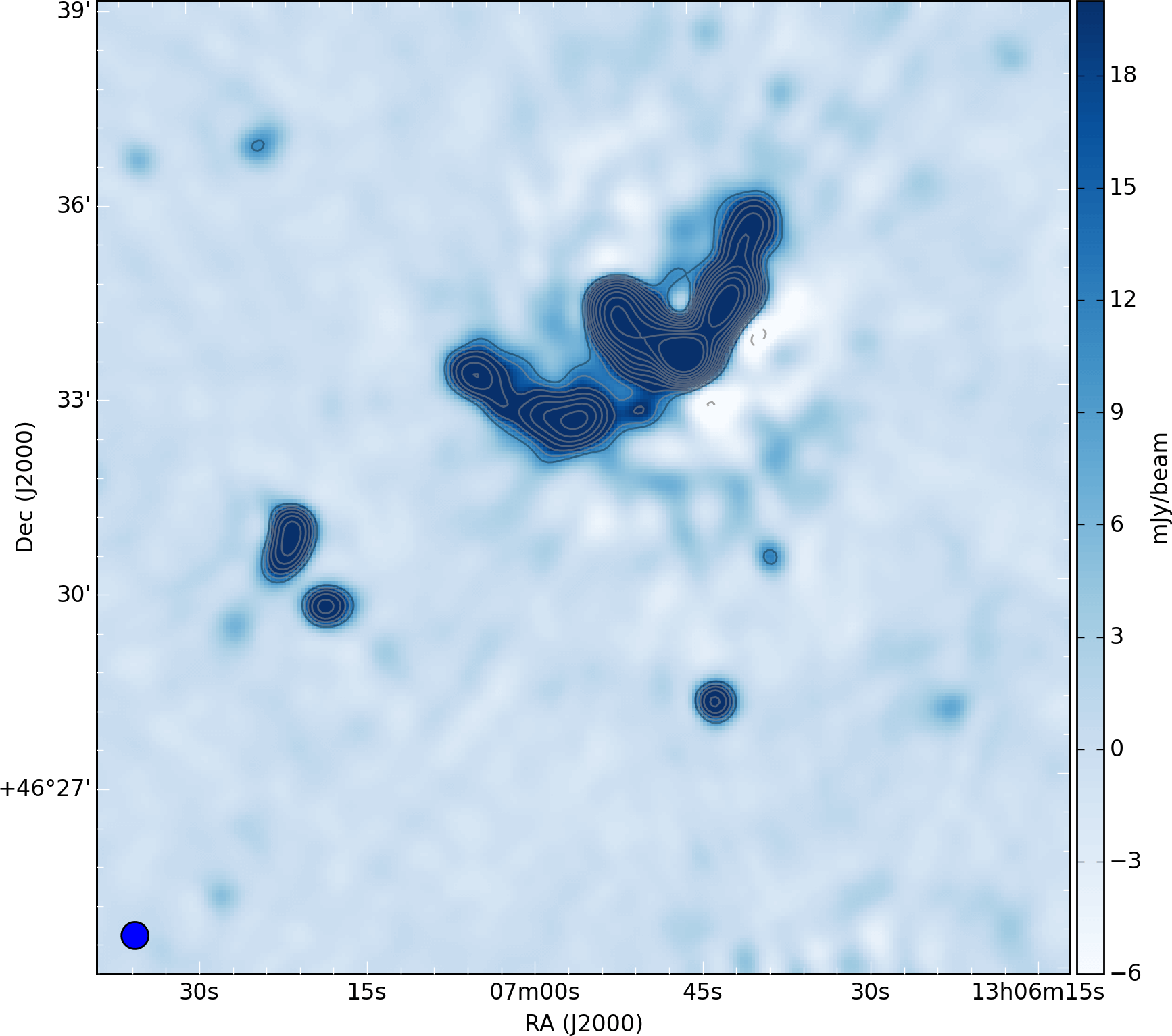} 
    \includegraphics[width=0.45\linewidth]{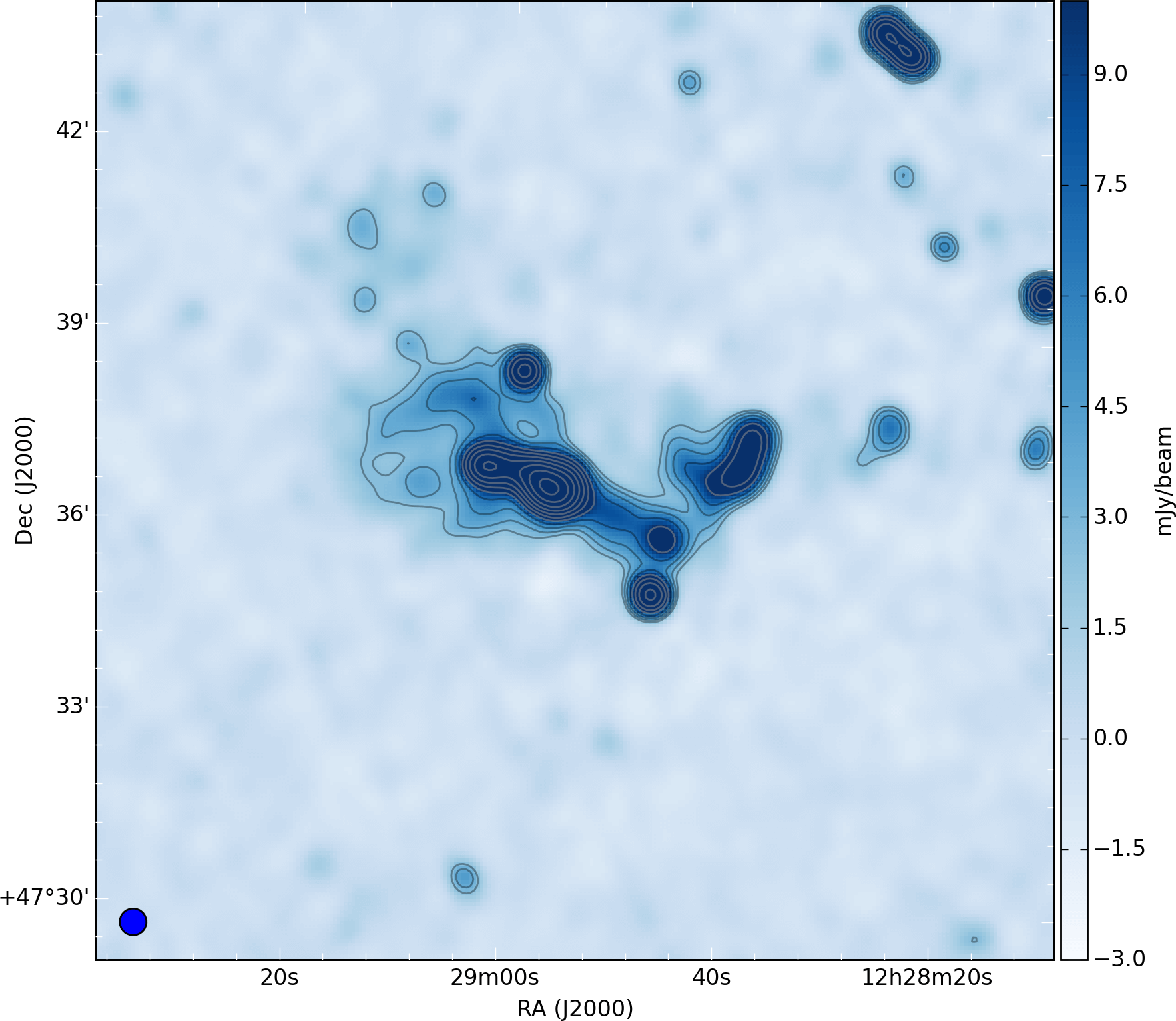}
   \caption{A sample of rare objects whose low-frequency emission we have characterised in the 25$\arcsec$ resolution preliminary data release images which are now available for download in FITS format. Clockwise from top left: the $z=6.21$ quasar J1429+5447 (circled); the nearby galaxies M\,106 and M\,51; the merging galaxy clusters Abell 1550 and Abell 1682; the Mpc-scale tailed radio galaxy IC 711 that resides in the galaxy cluster Abell 1314. The contours and colour scale vary between images and were chosen to best emphasise the structure of each object. The synthesised beam is shown in the bottom left corner of each image.}
   \label{fig:science-examples}
\end{figure*}

\section{Summary}
\label{sec:summary}

In this publication we have described the LoTSS, for which we aim to produce high fidelity images of the entire Northern sky with a resolution of $\approx5\arcsec$ and sensitivity of $\approx100\,\mu$Jy/beam at most declinations. We have summarised a survey strategy that should allow us to reach these ambitious observational aims. This consists of 3170 pointings each observed for 8\,hrs with frequency coverage from 120 to 168\,MHz and archived with sufficient spectral and time resolution to allow future spectral line studies and subarcsecond resolution imaging. The survey was initiated in mid-2014 in the region of the HETDEX Spring Field. As of November 2016 we will have observed 20\% of the sky above a declination of 25$^\circ$ and we are preparing to further increase our observing rate.

The main challenge of the survey is to robustly and efficiently perform a complex direction-dependent calibration of very large datasets. This is crucial in order to exploit the full potential of our LOFAR datasets. To demonstrate that such a reduction can reach our observational aims we have used the facet calibration technique, which was developed by \cite{vanWeeren_2015a} and \cite{Williams_2015}, to perform a direction-dependent calibration on one of the LoTSS datasets. The result is a high fidelity 120-168\,MHz image with a resolution of  $4.8\arcsec \times 7.9\arcsec$ and a sensitivity of 100\,$\mu$Jy/beam. These final high-resolution, high-fidelity LoTSS images will facilitate significant contributions to a wide variety of astronomical research areas and we intend to release such images to the wider scientific community in the future once our reduction strategy is finalised and we have processed a large area of the sky. In this publication we have instead publicly released preliminary images and catalogues from a completely automated  direction-independent calibration of 63 datasets in the region from right ascension 10h45m00s to 15h30m00s and declination 45$^\circ$00$\arcmin$00$\arcsec$ to 57$^\circ$00$\arcmin$00$\arcsec$. We have provided a brief summary of the scientific potential of these preliminary images and whilst they have lower fidelity, resolution, and sensitivity than those that we will make using a direction-dependent calibration strategy, they are still significantly more sensitive than those produced by any other existing large-area low-frequency survey and can allow for many scientific objectives of the LoTSS to be partially or completely realised (see e.g. \citealt{Brienza_2016}, \citealt{Harwood_2016}, \citealt{Heesen_2016}, \citealt{Mahony_2016}, \citealt{Shulevski_2015a} and \citealt{Shulevski_2015b} for examples). 

The images we have released cover an area of over 350 square degrees and contain over 44,000 radio sources when a detection threshold of seven times the noise is used. We have used a Monte-Carlo simulation to estimate that the catalogue is 90\% complete for sources with a flux density in excess of 3.9\,mJy/beam. Our astrometry checks of the catalogue reveal that the positional error is approximately 1.70$\arcsec$ and our photometry measurements indicated that our integrated flux density measurements are accurate to within 20\%.  

\section{Acknowledgements}

TS and HR acknowledge support from the ERC Advanced Investigator programme NewClusters 321271. PNB, JS, WLW, MJH and VH are grateful for support from the UK STFC via grants ST/M001229/1, ST/M001008/1 and ST/J001600/1. EKM acknowledges support from the Australian Research Council Centre of Excellence for All-sky Astrophysics (CAASTRO), through project number CE110001020. AD acknowledges support from the BMBF, through project 05A15STA. MH acknowledges financial support by the DFG through the Forschergruppe 1254. RM gratefully acknowledge support from the European Research Council under the European Union's Seventh Framework Programme (FP/2007-2013) /ERC Advanced Grant RADIOLIFE-320745. GJW gratefully acknowledges support from The Leverhulme Trust.  JZ gratefully acknowledges a South Africa National Research Foundation Square Kilometre Array Research Fellowship. 

LOFAR, the Low Frequency Array designed and constructed by ASTRON, has facilities in several countries, that are owned by various parties (each with their own funding sources), and that are collectively operated by the International LOFAR Telescope (ILT) foundation under a joint scientific policy. The National Radio Astronomy Observatory is a facility of the National Science Foundation operated under cooperative agreement by Associated Universities, Inc. Part of this work was carried out on the Dutch national e-infrastructure with the support of SURF Cooperative through grant e-infra 160022. We gratefully acknowledge support by N. Danezi (SURFsara) and C. Schrijvers (SURFsara).

\label{lastpage}
\end{document}